\documentclass[prb,aps,onecolumn,showpacs,floats,a4paper,times,pdf]{revtex4}
\usepackage{ulem}
\usepackage{cancel}
\usepackage{epsfig,bm,epsf,graphics}
\usepackage{xcolor}

\begin{document}
\title{DYNAMICS OF THE PRICE BEHAVIOR IN STOCK MARKETS: A STATISTICAL PHYSICS APPROACH}
\author{Hung T. Diep $^{1}$\footnote{corresponding author} and Gabriel Desgranges $^{2}$}
\address{$^{1}$ \quad Laboratoire de Physique Th\'eorique et Mod\'elisation,
Universit\'e de Cergy-Pontoise, CNRS, UMR 8089, 2 Avenue Adolphe
Chauvin, 95302 Cergy-Pontoise, Cedex, France ; diep@u-cergy.fr\\
$^{2}$ \quad Th\'eorie \'Economique, Mod\'elisation et Applications,
Universit\'e de Cergy-Pontoise, CNRS, UMR 8184,
33 Boulevard du Port, 95011 Cergy-Pontoise Cedex, France ; gabriel.desgranges@u-cergy.fr.}

\date{\today}

\begin{abstract}
We study the time evolution of stock markets using a statistical physics approach.  We consider an ensemble of agents who sell or buy a good according to several factors acting on them: the majority of the neighbors, the market atmosphere, the variation of the price and some specific measure applied at a given time. Each agent is represented by a spin having a number of discrete states $q$ or continuous states, describing the tendency of the agent for buying or selling.  The market atmosphere is represented by a parameter $T$ which plays the role of the temperature in physics: low $T$ corresponds to a calm market, high $T$ to a turbulent one. We show that there is a critical value of $T$, say $T_c$, where strong fluctuations between individual states lead to a disordered situation in which there is no majority: the numbers of sellers and buyers are equal, namely the market clearing.
The specific measure, by the government or by an economic organization, is parameterized by $H$ applied on the market at the time $t_1$ and removed at the time $t_2$.
We have used Monte Carlo simulations to study the time evolution of the price as functions of those parameters.
In particular we show that the price strongly fluctuates near $T_c$ and there
exists a critical value $H_c$ above which the boosting effect remains after $H$ is removed. Our model replicates the stylized facts in finance (time-independent price variation), volatility clustering (time-dependent dynamics) and persistent effect of a temporary shock.
The  second part of the paper deals with the price variation using a time-dependent mean-field theory. By supposing that the sellers and the buyers belong to two distinct communities with their characteristics different in both intra-group and inter-group interactions, we find the price oscillation with time.  Results are shown and discussed.
\end{abstract}


\maketitle


\section{Introduction}

\noindent Statistical physics models treat large ensembles of particles interacting with each other via interactions of various kinds. Much has been achieved in the understanding of their properties in different situations \cite{DiepSP}. In particular, collective behaviors such as those observed in phase transitions and in correlated dynamics have been demonstrated as consequences of the microscopic underlying interactions between particles, the space dimension and the system symmetry.\\

\noindent Social systems with a large number of members interacting with each other via some rules are not always well defined due to the very nature of human beings. Nevertheless, we can see macroscopic effects of a society via surveys, and behavior studies.  This means that in spite of the fact that a human is not a particle, collective behaviors retain only common features, washing away individual particularities in the averaging process.\\

\noindent The interaction of an individual with his neighbors under the social atmosphere (peace or unrest) has been used in many domains such as politics and sociology \cite{Galam1,Galam2,Castellano,Schweitzer,Oberschall,Bernstein}.
Quantitative sociodynamics using stochastic methods and models of social interaction processes has been investigated \cite{Helbing}. The validity of statistical laws in physics and social sciences has been examined \cite{Majorana}.
In particular, the role of the interaction network in the emergence of diversity of behavior has been shown \cite{Godoy}. Using interacting social networks, social conflicts have been studied using models of statistical physics \cite{Diep2017,Kaufman3,Diep2019} with the mean-field theory and Monte Carlo (MC) simulations.  It is noted that MC simulations have been used with success to study social phenomena \cite{Stauffer}.\\

\noindent We note that sociophysics models try to outline essential ingredients which govern macroscopic behavior of a society. Knowing these, one can predict a collective behavior in order to avoid undesired effects or to intervene to change the course of its evolution if necessary \cite{Bernstein}. We will see an example in this paper.\\

\noindent Statistical physics used to study economic problems has been called "econophysics" \cite{Stanley}. As in sociophysics, the correspondence between physical quantities and parameters can be interpreted in economic terms. We will give some discussion in the next section on this point.\\

Econophysics is a research field using theories and methods developed by physicists in order to solve problems in economics. These methods usually include uncertainty or stochastic processes and nonlinear dynamics which are inherent to economic matters \cite{Chakrabarti,Sornette,Bouchaud}.

\noindent In the present work, we consider the problem of the evolution during the time of the price in a commodity market by examining the effects of several parameters.  There is an enormous number of works dealing with the stock markets. We can mention a few of them: a book on the price theory and applications \cite{Hirshleifer,Hirshleifer1}, stock market crashes \cite{Sornette}, pricing policy \cite{Bouchaud}, price fluctuations \cite{Liu}, stock market behaviors \cite{Scalas,Shapira,Zunino} and scaling behavior in the dynamics of an economic index \cite{Mantegna}.\\

\noindent Price variation is subject of many investigations trying to find rules to predict or to understand the market dynamics. These investigations used many methods. Let us mention a few of them. Hsu et al. \cite{Hsu} have used a probability model for describing rates of return based on the hypothesized phenomenon of a changing variance,
Castanias  \cite{Castanias} has studied the correlation between macroinformation and the variability of stock market prices, Cutler et al. \cite{Cutler} have investigated some causes which move the stock prices by estimating the fraction of the variance in aggregate stock returns that can be attributed to various kinds of news. Bak et al. \cite{Bak} have constructed simple models of a stock market, and argued that the large variations may be due to a crowd effect, where agents imitate each other's behavior. Tse et al. \cite{Tse} have constructed complex networks to study correlations between the closing prices for all US stocks that were traded over two periods of time (from July 2005 to August 2007; and from June 2007 to May 2009).  Networks have been used in a work by Huang et al. \cite{Huang} who employed a threshold method to construct China's stock correlation network and then study the network's structural properties and topological stability. They found that it follows a power-law model.  Sun et al. \cite{Sun} have used a multifractal analysis of Hang Seng index to study the variation of the Hong Kong stock market, Chang et al. \cite{Chang} have used a fuzzy model to test with success on the Taiwan Electronic Shares from the Taiwan Stock Exchange. Song et al. \cite{Song} have investigated the evolution of worldwide stock markets using correlation structure and correlation-based graphs.\\

\noindent T. Lux has given a review \cite{Lux} on stochastic models borrowed from statistical physics using microscopic interactions between a larger number of traders to deduce macroscopic regularities of the market, independent of microscopic details. We will show in the present work the macroscopic behavior after the time averaging does not depend on particularities of each agent but on a set of general parameters  such as interaction strength between them and the economic environment.  On the same line, R. Cont \cite{Cont} has shown various statistical properties of asset returns with emphasis on properties common to a wide variety of
markets and instruments. This analysis invalidates many of the common statistical approaches used to study financial data sets mentioned above \cite{Hsu,Castanias,Cutler,Bak,Tse,Huang,Sun,Chang,Song}. We need therefore a general model which does not use empirical rules in the course of calculation. This motivates our present work.\\

\noindent The above mentioned works show that there have been many models trying to find rules for price variation in stock markets. Large variations in stock prices happened so frequently that one can raise doubts about universal models which can be applied to many situations.\\

\noindent In the following, we introduce a new model inspired from models of statistical physics \cite{DiepSP} but with modifications where they should be for econophysics. Our model contributes to the family of  already abundant stock price variation models, but as seen below,  it gives rise to new features not seen before in the price variation. Our model uses an assembly of agents interacting with each other and with the economic temperature as well as with economic measures taken by the government. This model is studied using Monte Carlo simulations and by the time-dependence mean-field theory.\\

\noindent The aim of the paper is to study the variation of the price of a good under the effects of the interaction strength between agents, the economic environment (economic crisis, economic agitation, ...) and boosting measures taken by the government.\\

\noindent We note that the change of price directly yields gross returns. From the price variation, one can deduce gross returns (namely before fees, taxes, inflation, ...) as seen below. Our objective is to show that the price dynamics which can be used to deduce the stylized facts. The first one is the case where the price of time $t$ is independent of that at time $(t-1)$,  real-time fluctuations yields  the time average over a lapse of time independent of the time.  This stylized fact is described by our model as seen below. The second stylized fact concerns autocorrelated  fluctuations such as volatility clustering explained below. This is found in our model as well (see section \ref{MCdiscussion}).\\

\noindent As said above, a stylized fact in finance is the so-called "volatility clustering". Volatility clustering was suggested by Mandelbrot in 1963 from the observation that "large changes tend to be followed by large changes, of either sign, and small changes tend to be followed by small changes."\cite{Mandelbrot} Many works, mostly empirical, have been done since then \cite{Ding,Ding1,Ding2,Cont1,Barndorff} to study the persistence of some tendencies due to an autocorrelation, namely variation at time $t$ depends on the variation in the past.  This is one of the objective of the present paper.  We will show that a specific measure taken by the government or an economic organization to boost or to lower the market price can have a long-lasting effect.
The absence of a model to explain why  a temporary shock can generate persistent effects has motivated our present work. The reader is referred to textbooks on macroeconomics for this point (see for example Refs. \onlinecite{book1,book2}). Our model presented below explains this stylized fact.\\

\noindent This paper is organized as follows. Section \ref{model} is devoted to the description of our model with emphasis put on the meaning of physical parameters in economic issues. The method of Monte Carlo simulation is described in this section. In section \ref{MCresults}, we present the results of three models: the 3-state model, the 5-state model and the continuous model.
Section \ref{MFT} is devoted to a mean-field study. The model used in this section uses the assumption that the sellers and the buyers belong to two communities with different intra-group interactions and different inter-group interactions. This non-symmetric interactions give rise to the price oscillations in a region of market temperature. Mean-field results are shown and discussed.     Concluding remarks are given in section \ref{concl}.

\section{Model and Method}\label{model}

\subsection{Model}

\noindent We recall a simple textbook static model of a market for one commodity (one can say one
good).
There is a population of agents facing an aggregate supply function $F\left(
P\right) $.\ The decision of every agent $i$ is a demand of good $x_{i}$.\
The price $P$ is determined by market clearing ({\it i.e. }supply=demand)%
\begin{equation}\label{m1}
\sum_{i}x_{i}=F\left( P\right)
\end{equation}
Consider the inverse supply function $p\left( .\right) =F^{-1}\left(
.\right) $.\ Market clearing rewrites%

\begin{equation}\label{m2}
P=p\left( \sum_{i}x_{i}\right)
\end{equation}

\noindent We assume a linear and increasing function $p$%
\begin{equation}\label{m3}
p\left( X\right) =A+BX
\end{equation}
where $X= \sum_{i}x_{i}$, $A$ and $B$ are two positive real parameters.

\noindent The demand $x_{i}$ of agent $i$ can take a finite number of values.\ For
example, $x_{i}\in \left\{ -1,0,1\right\} $.

\bigskip

\noindent This static game is repeated indefinitely (time is discrete: $t=1,2,...$):
\begin{equation}\label{m4}
P(t)=p\left( \sum_{i}x_{i,t}\right)
\end{equation}

\noindent The decision $x_{i,t}$ of every agent $i$ are revised according to some
learning algorithm based on the past values $P(t-1)$ and $x_{j,t-1}$ where
the agents $j$ are $i$'s neighbors.
This algorithm has 2\ main features:\\

- $x_{i,t}$ is increasing in the $x_{j,t-1}$ (agent $i$ imitates his
neighbors' past behavior)

- the effect of $P(t-1)$ on $x_{i,t}$ is twofold: a higher price decreases $i$'s demand (direct effect), a higher price reveals that other agents buy a
lot, which increases $i$'s demand (indirect informative effect, due to $i$'s
imitation of others' behavior). Hence, $x_{i,t}$ can be either increasing or
decreasing in $P(t-1)$. Since $P(t-1)=p\left( \sum_{i}x_{i,t-1}\right) $,
the demand $x_{i,t}$ can be increasing or decreasing in $\sum_{i}x_{i,t-1}$.

\bigskip

\noindent The practical implication of these remarks for numerical simulations is that
the two cases ($x_{i,t}$ increasing or decreasing in $\sum_{i}x_{i,t-1}$)
should be considered.\\

\noindent In the simulation, we translate the above rule into a model in statistical physics using a spin language.
Each agent has several states which can be represented by a spin. A spin in statistical physics is an object characterized by a number of internal states. For example, an Ising spin has two states up and down, a $XY$ spin has two continuous components, a Heisenberg spin has three components. In this paper we take a spin with an integer amplitude $S$ but, unlike the Ising spin, it has $2S+1$ states: -$S$, $-S+1$, $-S+2$,..., 0, 1, 2,..., $S$. For $S=1$, one has thus three states $M=-1,0,1$.
In econophysics, a spin state is an economic action of an agent: selling, buying or waiting, for example. A spin $\sigma_i$ describes the state of an agent $i$. $\sigma_i$ has several states, for example $\sigma_i=-1,0,1$. Each state represents an action: let's define $\sigma_i=1$ for buying action, $\sigma_i=-1$ for selling, and $\sigma_i=0$ for waiting.  The action of agent $i$ is the result of different mechanisms described in what follows.  If $S=2$, each agent has $2S+1=5$ states from -2S to 2S, namely -2, -1, 0, 1, 2, expressing  respectively a strong desire to sell (-2), a moderate desire to sell (-1), waiting (0), a moderate desire to buy (+1) and a strong desire to buy (+2). Therefore, the more states the more degrees of will. \\

\noindent Let us discuss now the correspondence between parameters from statistical physics and their meanings in economics.  First, the temperature $T$ in physics is an external parameter which acts on each particle in a way to make it disordered (thermal agitation): higher $T$ higher disorder. In econophysics, the market temperature $T$ expresses the economic atmosphere resulting from many factors such as political situation, economic crisis and international conflicts. In econophysics, low $T$ means stability, high $T$ means unstable situation.  Second, agents interact with each other via an imitation interaction $J$ which leads to some order or collective structure, in contrast with $T$ which favors disorder.  Therefore, there is a competition between $T$ and $J$.\\

\noindent We consider an ensemble of $N$ agents.
We suppose that the energy associated with agent $i$ is given by

\begin{equation}\label{energy}
E_i (t)=\sigma_i(t)\  [-J\sum_j \sigma_j(t-1)] +a\ \sigma_i(t)\ [N(\mbox{up},t-1)-N(\mbox{down},t-1)]-H\ \sigma_i(t)
\end{equation}
where $a > 0$.  Let us explain each term of the above equation:\\

i) The first term represents the sum of the influence on $\sigma_i(t)$ at the time $t$ by his neighbors' attitudes at the time $(t-1)$. For simplicity, we assume here all neighbors have the same interaction $J$ with $\sigma_i(t)$. This will not alter general aspects of the model. The agent imitates the majority of his neighbors\\

ii) In addition to the influence of neighbors given by the first term,  $E_i(t)$ also depends on the price tendency given by the second term: let $N(\mbox{up},t-1)$ ($N(\mbox{down},t-1)$) be the number of the people who wants to buy (sell) at the previous time $t-1$.  The price is proportional to $N(\mbox{up},t-1)-N(\mbox{down},t-1)$, namely\\

* if $N(\mbox{up},t-1) > N(\mbox{down},t-1)$, i. e. more people who buy, so the price is high (increasing), $\sigma_i$ may take the value $-1$ (sell), against the buying tendency of the first term, to take benefits of selling at a high price\\

* if $N(\mbox{up},t-1) <  N(\mbox{down},t-1)$, i. e. more people who sell, so the price is low (decreasing), $\sigma_i$ may take the value $+1$ (buy), against the imitation tendency of the first term, to take advantage of buying at a low price\\

iii) The third term is a market-oriented measure to boost buying if $H$ is positive, or to favor selling if $H$ is negative.  This measure can be applied for a lapse of time and is removed to leave the market evolve.\\

\noindent Note that the decision of $\sigma_i(t)$ at a given $T$ (to buy, to sell or to wait) depends on the balance of the three terms in Eq. (\ref{energy}). It is the total sum that matters.\\

\subsection{Monte Carlo Method}

\noindent Let us now briefly describe the Monte Carlo algorithm used in this work (see details in Refs. \onlinecite{Diep2017,Diep2019,DiepTM}): (i) We generate a system of $N$ individuals. Each of them has a random initial value among -1, 0 and 1, with a chosen percentage of each kind, (ii) We fix the ``temperature" $T$ which represents the market agitation, analogous to the temperature in physics: low (high) $T$ corresponds to peaceful (agitated) economic environment, (iii)  At the time $t$ we calculate $E_i (t)$ of the individual $\sigma_i(t)$, using Eq. (\ref{energy}) with the states of its neighbors at the previous time ($t-1$), (iv) We take randomly a new value of $\sigma_i(t)$ and calculate the energy difference $\Delta E_i=E_i(\mbox{new})-E_i(t)$ between the new and the old states and we apply the Metropolis algorithm \cite{DiepTM} to update the spin state.
  We take then another $\sigma_i(t)$ and repeat the updating procedure described above until the states of all agents are updated.  We say we achieve one Monte Carlo sweep (or step), (v) We calculate $N(\mbox{up},t-1)=\sum_i \delta(\sigma_i-1)$ and $N(\mbox{down},t-1)=\sum_i \delta(\sigma_i+1)$ and memorize all $\sigma_i(t)$,
We start again another Monte Carlo sweep (namely for $t+1$) by repeating the updating procedure from point (iii).  We do that for a large number of times to follow the time evolution of the system.\\

%

\noindent We note that the price  $P(t)$ can be defined as
\begin{equation}\label{price}
P(t) = a\ [N(\mbox{up},t)-N(\mbox{down},t)]+A
\end{equation}
where $a$ is a proportional constant and $A$ denotes the stable price determined by the market clearing, namely  when $N(\mbox{up},t)=N(\mbox{down},t)$.
The gross return is defined as

\begin{equation}\label{return}
R(t) = [P(t)-P(t-1)]/P(t-1)
\end{equation}

\noindent In practice, we can start the simulation at a given $T$ with different initial conditions, for example with 60\% of $\sigma$ up (percentage of buyers), 40\% of $\sigma$ down (percentage of sellers), and leave the system evolve to the final state at that given $T$.
We note that at the beginning of the market exchange the initial percentages of buyers and sellers may not correspond to the economic temperature $T$, so the variation of the price may be strong, and this lasts for a lapse of time depending on the initial percentages and $T$.  After this initial variation, the price fluctuates around a mean value. The first period can be said "time-dependent dynamics". The second period when the price fluctuates around a mean value is a time-independent dynamics and corresponds then to the stylized fact of price informational efficiency (namely, the dynamics of price is time-independent). Both kinds of dynamics occur in our simulations as seen below.

Our model is the Ising-like model which has been studied in details on many aspects in statistical physics. In particular, the dynamics of the spin-spin correlation (correlation function) which decays with distance, and of course the autocorrelation which is a function of time.  Unlike many earlier models in econophysics, our model takes into account the economic atmosphere characterized by the economic temperature $T$. Our model shows that there is a critical temperature $T_c$ above which the market clearance sets in. The Ising model shows that the relaxation time, or the correlation between members, depends strongly on $T$: in the region around $T_c$ (critical region), the system is in the regime of "critical slowing-down" where the relaxation time is extremely long $\tau \propto 1/|T-T_c|^{z\nu}$ where $z$ is the dynamical exponent and $\nu$ the correlation exponent ($z=2$, $\nu=0.63$ for 3D Ising model) \cite{Hohenberg}. This has been shown years ago . In our econophysics model, it is in this critical region that interesting phenomena happen, for instance
(i) when $T$ is far from $T_c$, the price fluctuations with time are very small (see results shown below). While when $T$ is close to $T_c$, price fluctuations are very strong (see results below), this may be called volatility clustering, namely strong variations induce strong variations;
(ii) the persistent effect occurs at $T$ just below $T_c$: the effect of  a strong enough boosting measure at $T<T_c$ will last very long (see results below);
(iii) Autocorrelation is known to be exponentially decayed to zero in the Ising model for $T>T_c$ \cite{Janke}. For $T<T_c$, the autocorrelation decays to a finite value, this is called a fat tail in econophysics but this is well known in the Ising model.  Note that in Ising disordered systems such as spin glasses, frustrated systems or topological textures (skyrmions, ...) the autocorrelation follows stretched exponential laws \cite{Phillips,Ogielski,Ngo2014,Sahbi2018}.

Note that the exponential decay of the autocorrelation and the stretched-exponential autocorrelation as well as fat-tailed empirical distribution of returns have been found in various models in econophysics. For reviews, the readers are referred to Refs. \cite{Chakraborti1,Chakraborti2}.

\section{Monte Carlo simulation results}\label{MCresults}

\noindent We perform Monte Carlo simulations using a network of $N=4\times 12^3=6912$ agents on a face-centered cubic lattice of linear dimension $L=12$ where each agent has 12 nearest neighbors (NN). We suppose that each agent interacts with his/her NN with the same interaction strength $J$ as indicated in Eq. (\ref{energy}).\\

\noindent It is interesting to examine first the stability of the market as a function of $T$. By stability we mean that the interaction term in Eq. (\ref{energy}) dominates, namely the collective effect with correlation among agents is present. This occurs when $T$ is lower than a value $T_c$ beyond which agents are independent of each other because the agitation of $T$ which breaks the correlation.  This temperature $T_c$ is called ``transition temperature" which is seen by anomalies in various quantities such as the system energy $E$, the order parameter $M$, fluctuations of $E$ and $M$ called calorific capacity $C_V$ and susceptibility $\chi$ in statistical physics.  These quantities are defined as

\begin{eqnarray}
E(t)&=&\sum_i E_i(t)\\
\langle E \rangle &=&\langle  E(t)\rangle \label{E}\\
C_V&=&\frac{\langle  E(t)^2\rangle -\langle  E\rangle^2}{k_BT^2}\label{CV}\\
\langle M \rangle &=&\frac{1}{N}\langle  M(t)\rangle =\frac{1}{N}\langle  \sum_i \sigma_i(t)\rangle \label{M}\\
\chi&=&\frac{\langle  M(t)^2\rangle -\langle  M(t)\rangle^2}{k_BT}\label{CHI}
\end{eqnarray}
where care should be taken in the sum in $E(t)$ is to avoid the double counting of the interactions in the first term of Eq. (\ref{energy}). $M$ is the ``magnetization" in statistical physics which represents the order parameter,  $\langle  ...\rangle$ indicates the thermal average, namely the average taken over all microscopic states at the temperature $T$ which occur during the simulation time. Note that $C_V$ measures the fluctuations of the energy $E$, and $\chi$ measures the fluctuations of $M$. \\

\begin{figure}[h!]
\vspace{5pt}
\centering
\includegraphics[scale=0.20,angle=0]{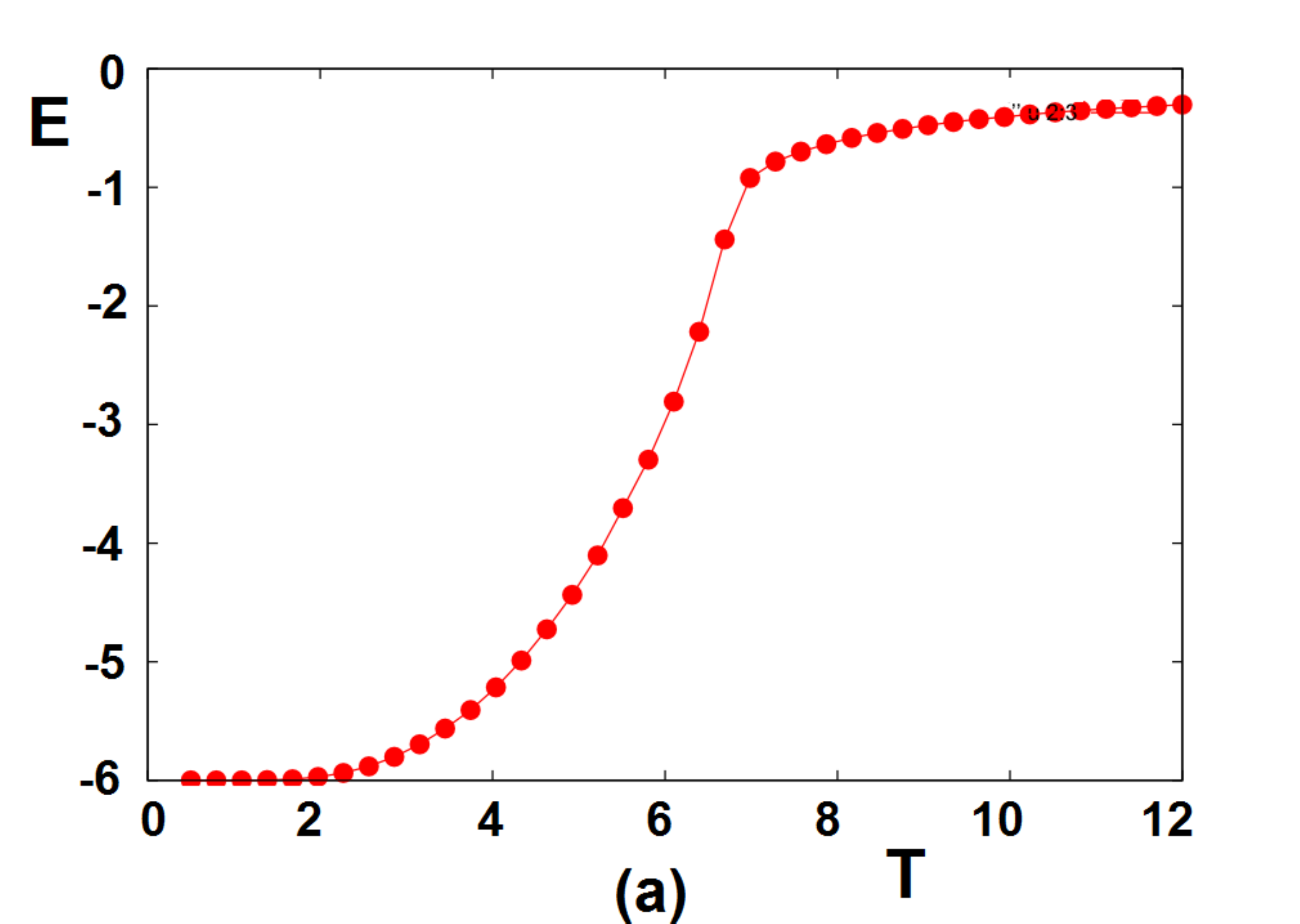}
\includegraphics[scale=0.20,angle=0]{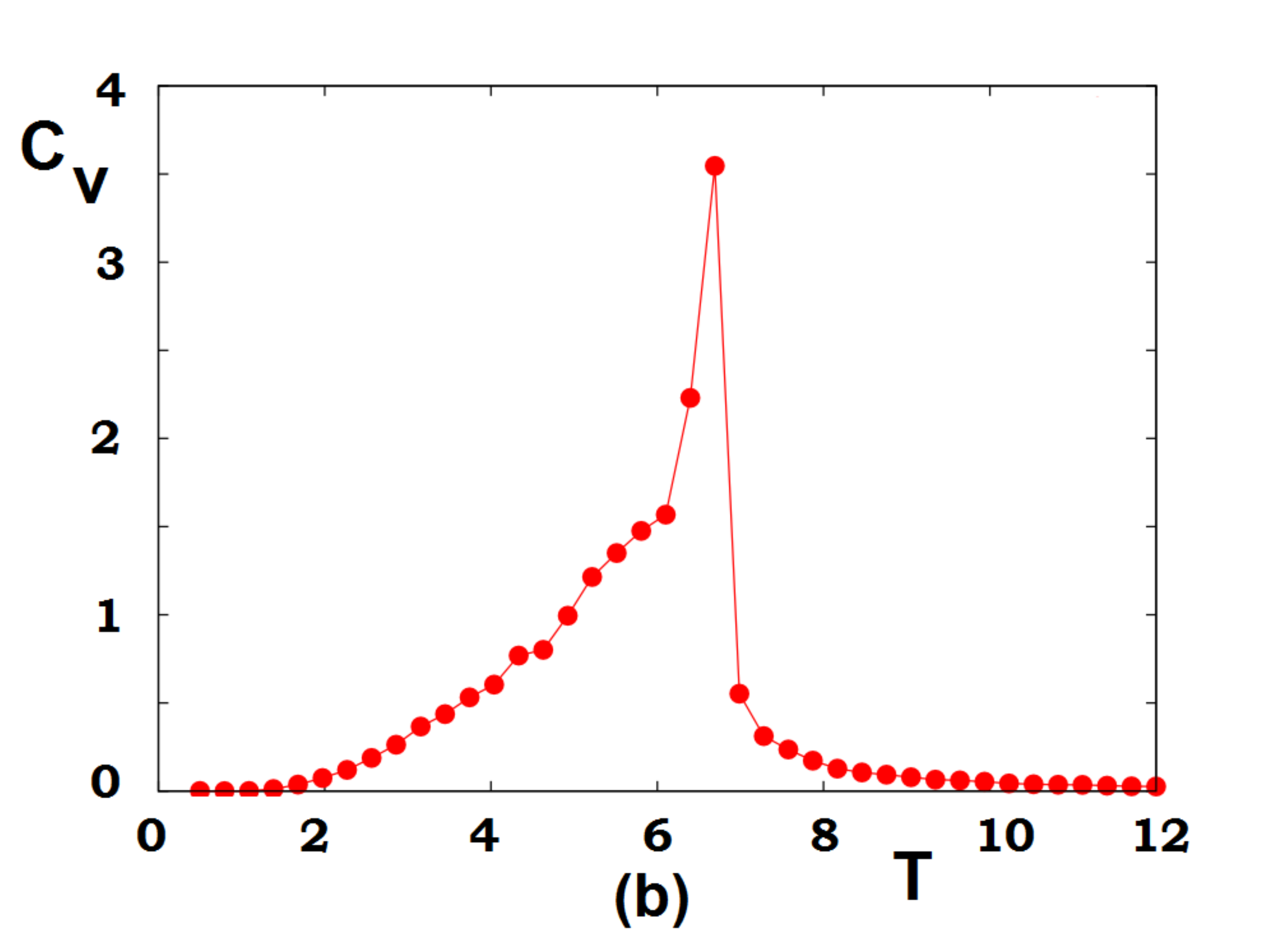}
\includegraphics[scale=0.20,angle=0]{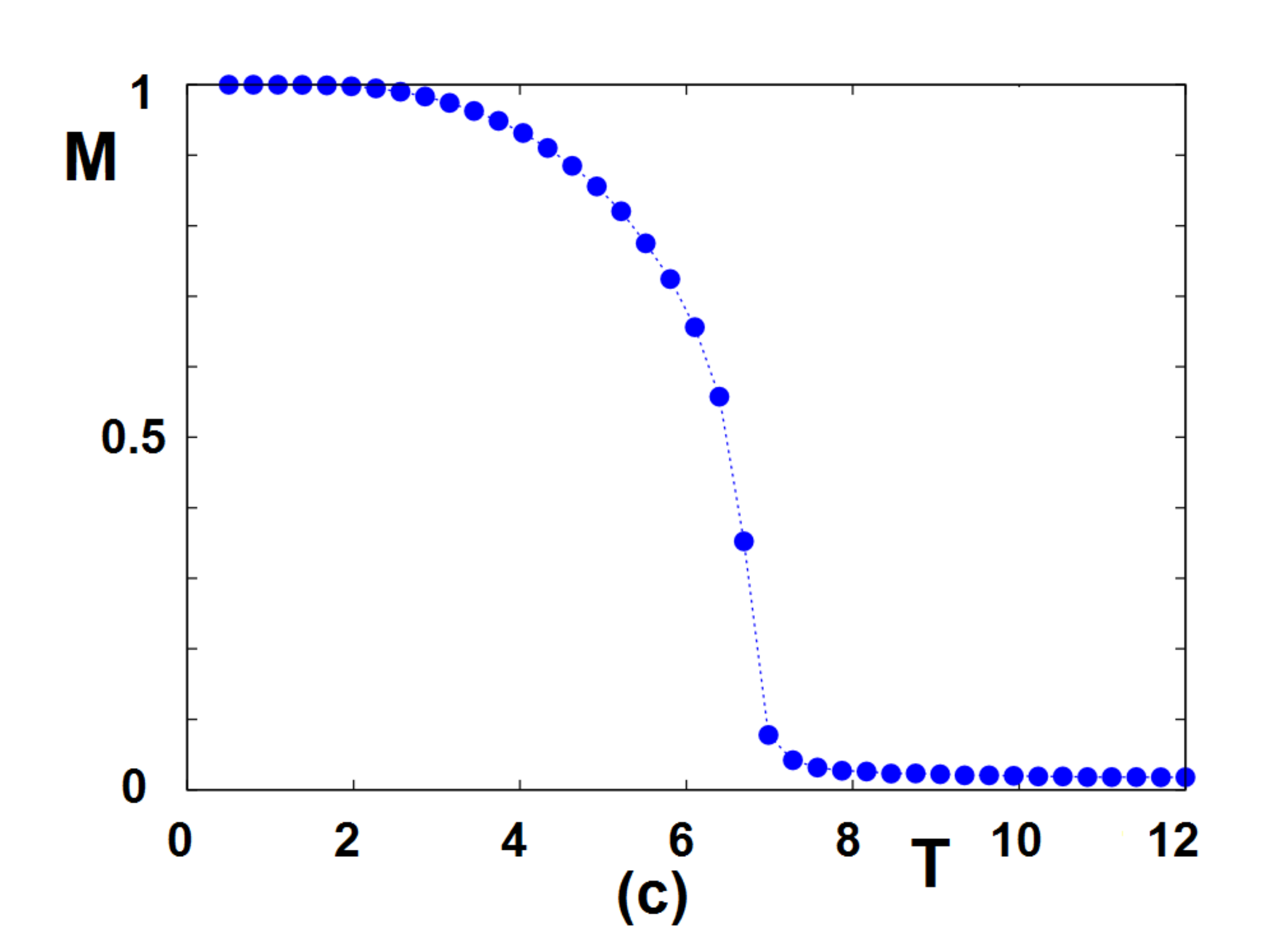}
\includegraphics[scale=0.20,angle=0]{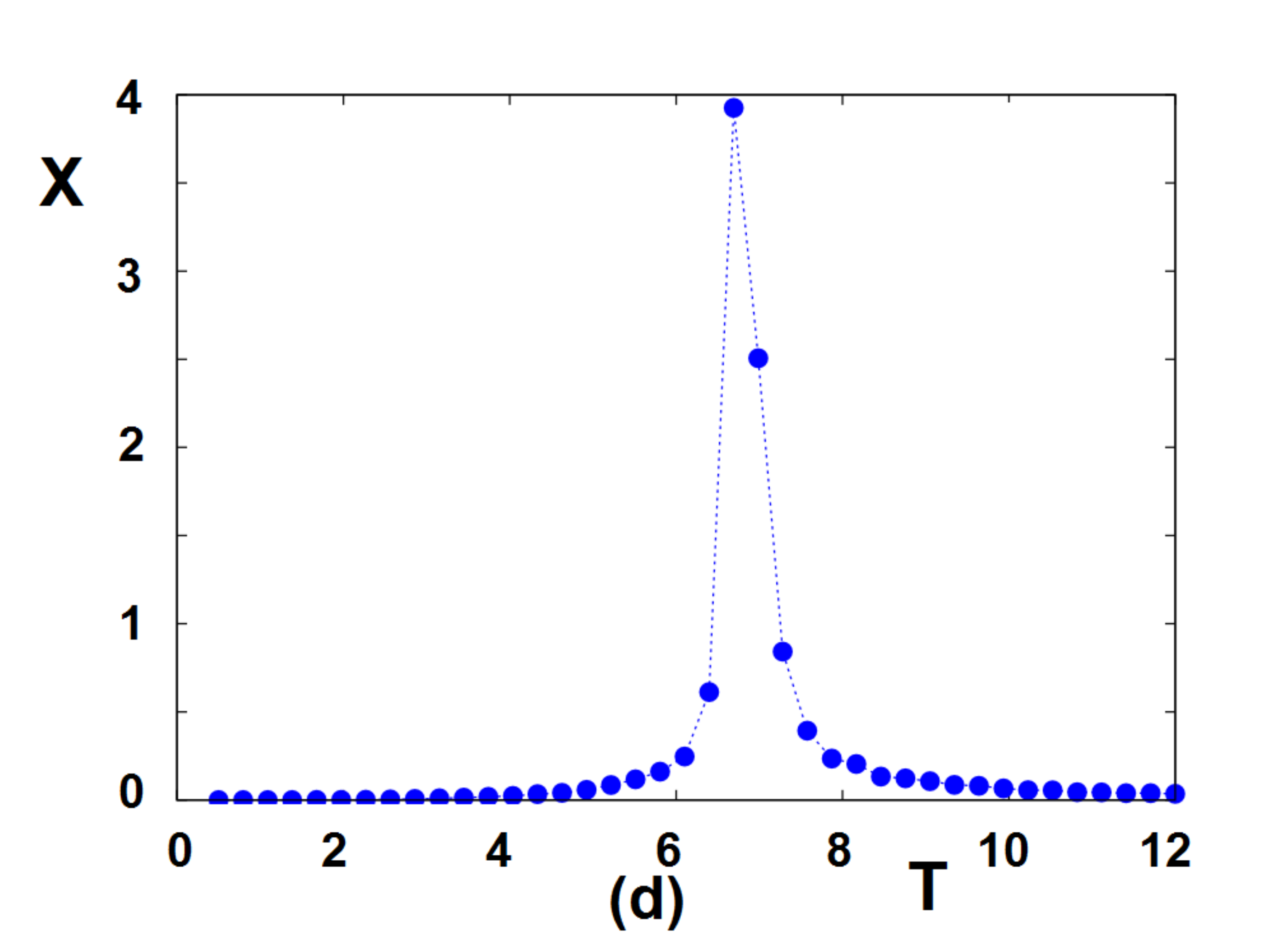}
\vspace{5pt}
\caption{(a) Energy per individual $E=\langle E \rangle /N$, (b) specific heat per individual $C_V$, (c) order parameter $M=\langle M \rangle $, (d) susceptibility $\chi$, versus $T$.  Parameters in Eq. (\ref{energy}): $J=1$, $a$=3, $H=0$.  See text for comments. }
\label{stab}
\end{figure}

\noindent We show in Fig. \ref{stab} the above quantities as functions of $T$, taking $J=1$, $a=3$ and $H=0$.
These quantities show the role of $T$ which allows us to choose appropriate temperature regions when examining the dynamics of the system.  As seen in the figure, the energy changes the curvature  and the order parameter falls to zero at $T_c=6.6957$ (the residue of $M$ after $T_c$ is due to the size effect). The fluctuations are very strong, $C_V$ and $\chi$ show a peak at $T_c$, indicating the transition from the low-$T$ phase to high-$T$ disordered phase.  We will see below that it is in the critical region just below $T_c$ that interesting dynamics occurs.\\

\noindent In the following, we show the fluctuations of the buyer and seller percentages as functions of time at various temperatures. The chosen values of $T$ for the figures do not have a particular importance, but it is important to indicate that they are below or above $T_c$, in the critical region or not, because as will be seen the price dynamics is interesting only in the critical region of temperature. Also, we have taken $J=1$ (unit of energy) and $a=3$ in the examples shown below, but varying $a$ around this value does not change the conclusion. The choice of $a$ is guided by the fact that we would like to have the competition between the first term and the second term in Eq. (\ref{energy}): the first term represents the interaction of an agent with his neighbors (local interaction) while the second term represents the interaction of an agent with the whole system (global interaction). $a$ is chosen so that the two terms have the same order of magnitude.   Of course if $J\gg a$, $J$ will dominate: an agent decides to sell or to buy according to states of his neighbors only, the influence of the market price is small. If $a$ is much larger than $J$, then the influence of his neighbors is neglected. The dynamics is not interesting. In systems with competing interactions  such as $J=1$ and $a=3$, the dynamics is non trivial as shown in Figs. 2-7.  However changing $a$ around this value will not change qualitatively the results. Note that when we change the number of individual states from 3 (section 3.1) to 5 (section 3.3) or to infinity (section 3.4) the value of $a$ changes in order to maintain a competition with the $J$ term. We took $a=6$ for the case of 5 states. Again, varying $a$ around this value will not change the qualitative features of the price dynamics.

\subsection{Case of more people who sell}
\begin{figure}[h!]
\vspace{5pt}
\centering
\includegraphics[scale=0.20,angle=0]{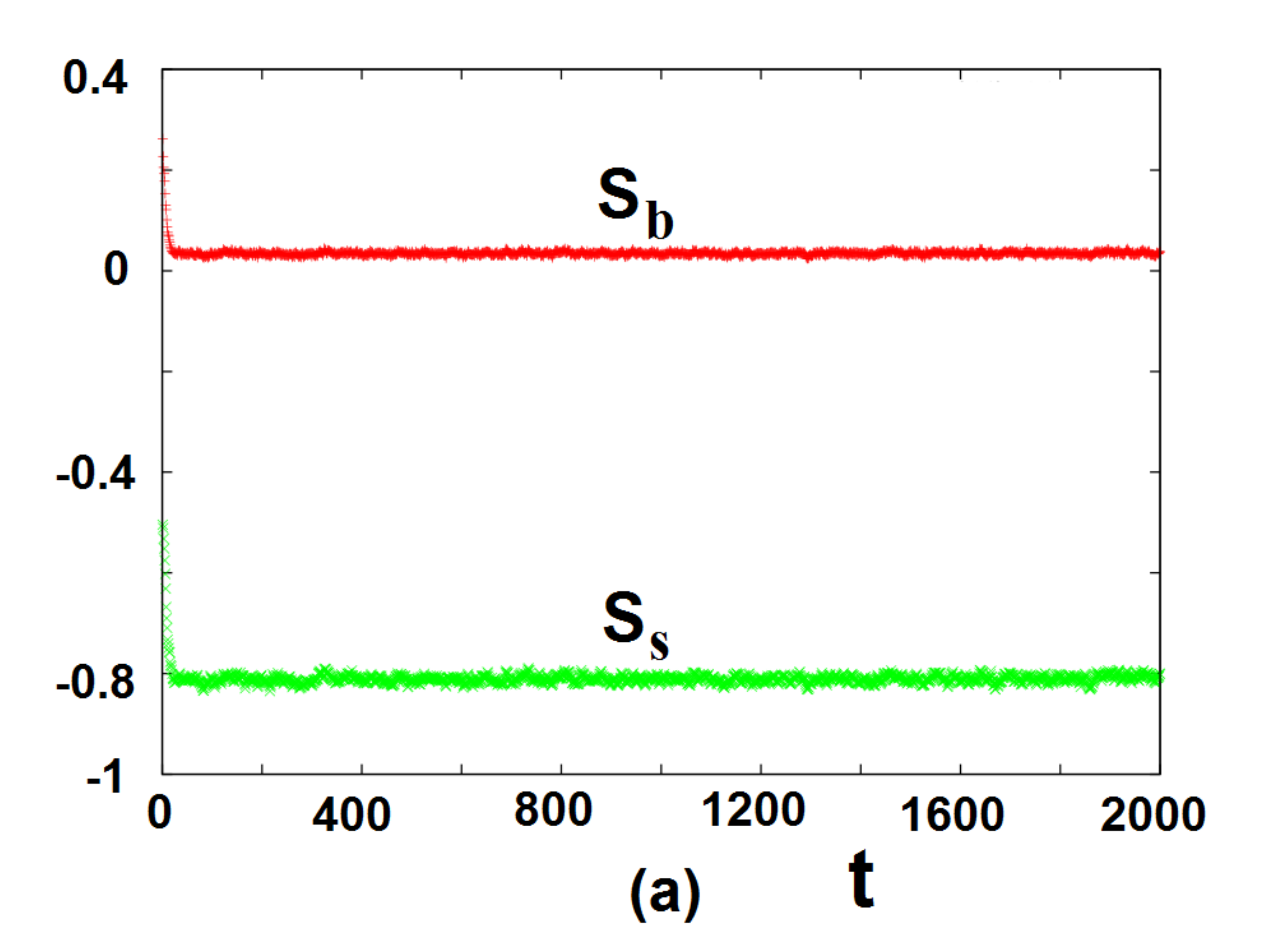}
\includegraphics[scale=0.20,angle=0]{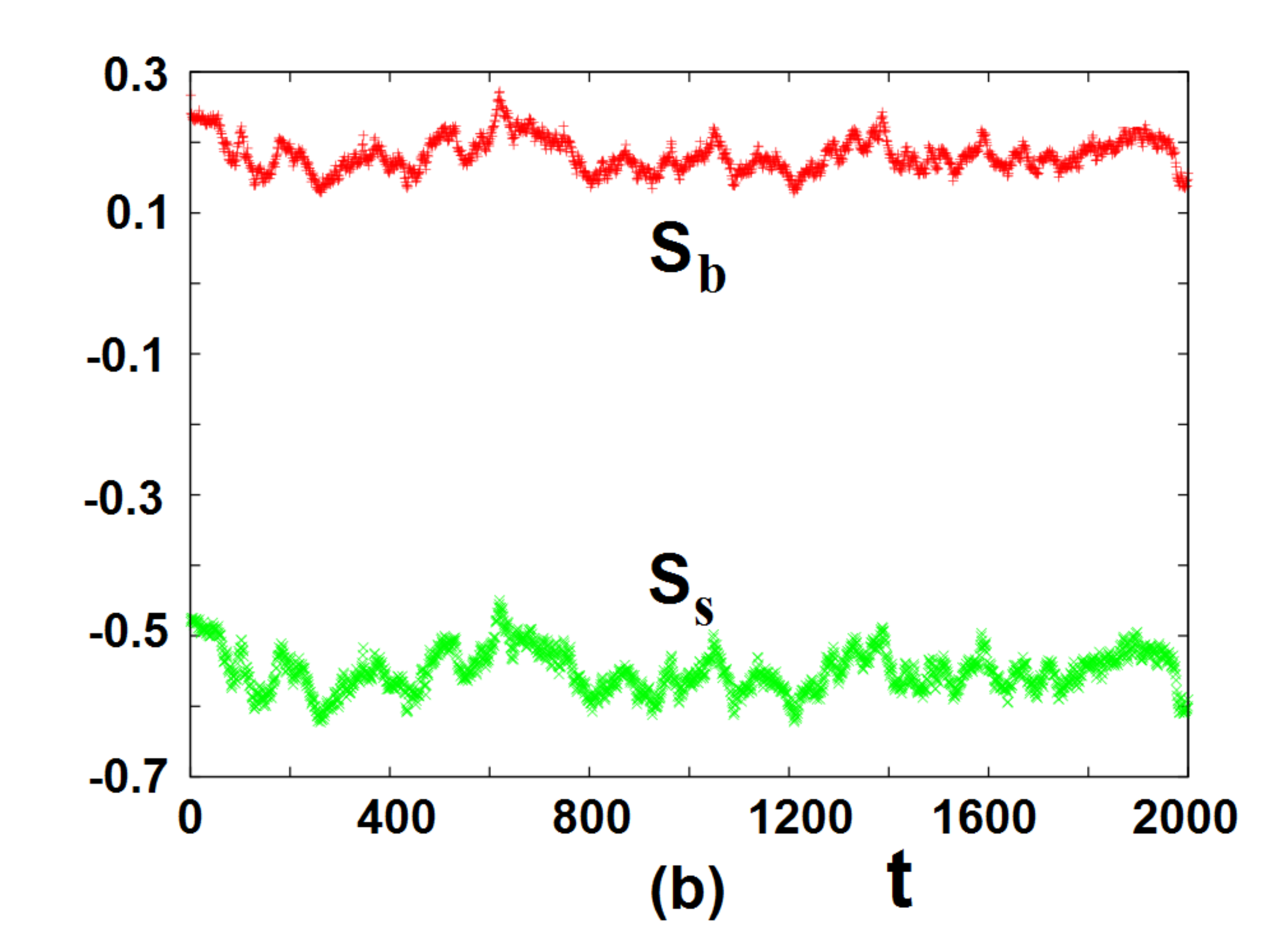}
\vspace{5pt}
\caption{Time evolution of the percentages $S_b$ and $S_s$ of  buyers (red) and sellers (green)   at  temperatures (a) $T=5.513$ (well below $T_c$) and (b) $T=6.692$ (just below $T_c$).
Initial condition: 60\% of agents who sell, 40\% of
agents who buy. Parameters used: $J=1$, $H=0$.  See text for comments. }
\label{ffig2}
\end{figure}

\noindent We show in Fig. \ref{ffig2} the time evolution of the seller and buyer percentages, $S_s$ and $S_b$ respectively, at some typical market temperature $T$ for $J=1$, $a=3$ and $H=0$. We use the initial condition: more people to sell than people to buy. At a low $T$, each agent imitates the majority of his surrounding neighbors who sells so that the price decreases and the system quickly comes to an equilibrium of percentages at that given $T$, as seen in Fig. \ref{ffig2}a. Near $T_c$, the difference in the percentages is reduced as seen in Fig. \ref{ffig2}b. Note the strong fluctuations of $S_s$ and $S_b$ which occur in the critical region of market temperature. For $T > T_c$ the two populations are equal (not shown), as expected.

\begin{figure}[h!]
\vspace{5pt}
\centering
\includegraphics[scale=0.20,angle=0]{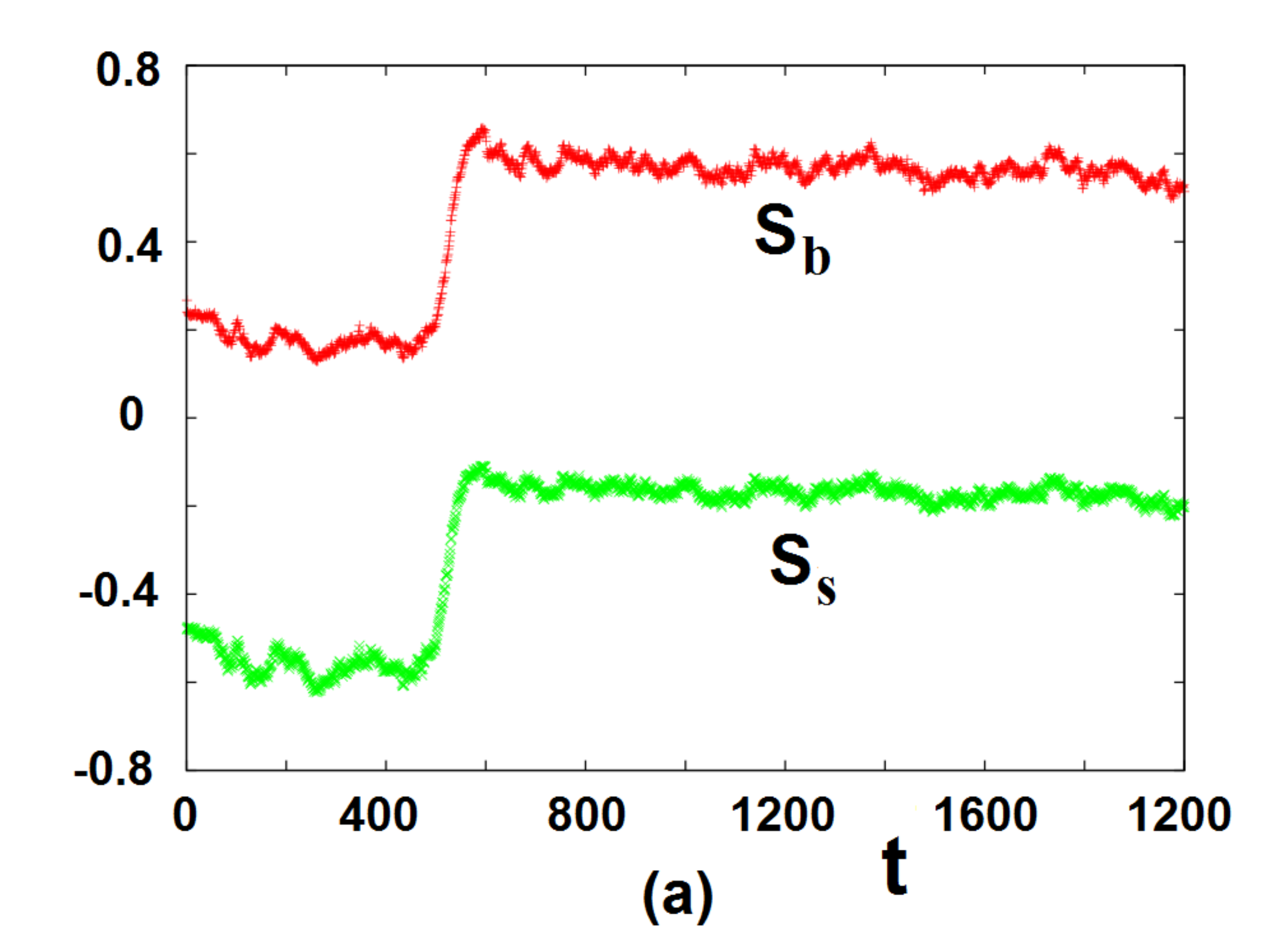}
\includegraphics[scale=0.20,angle=0]{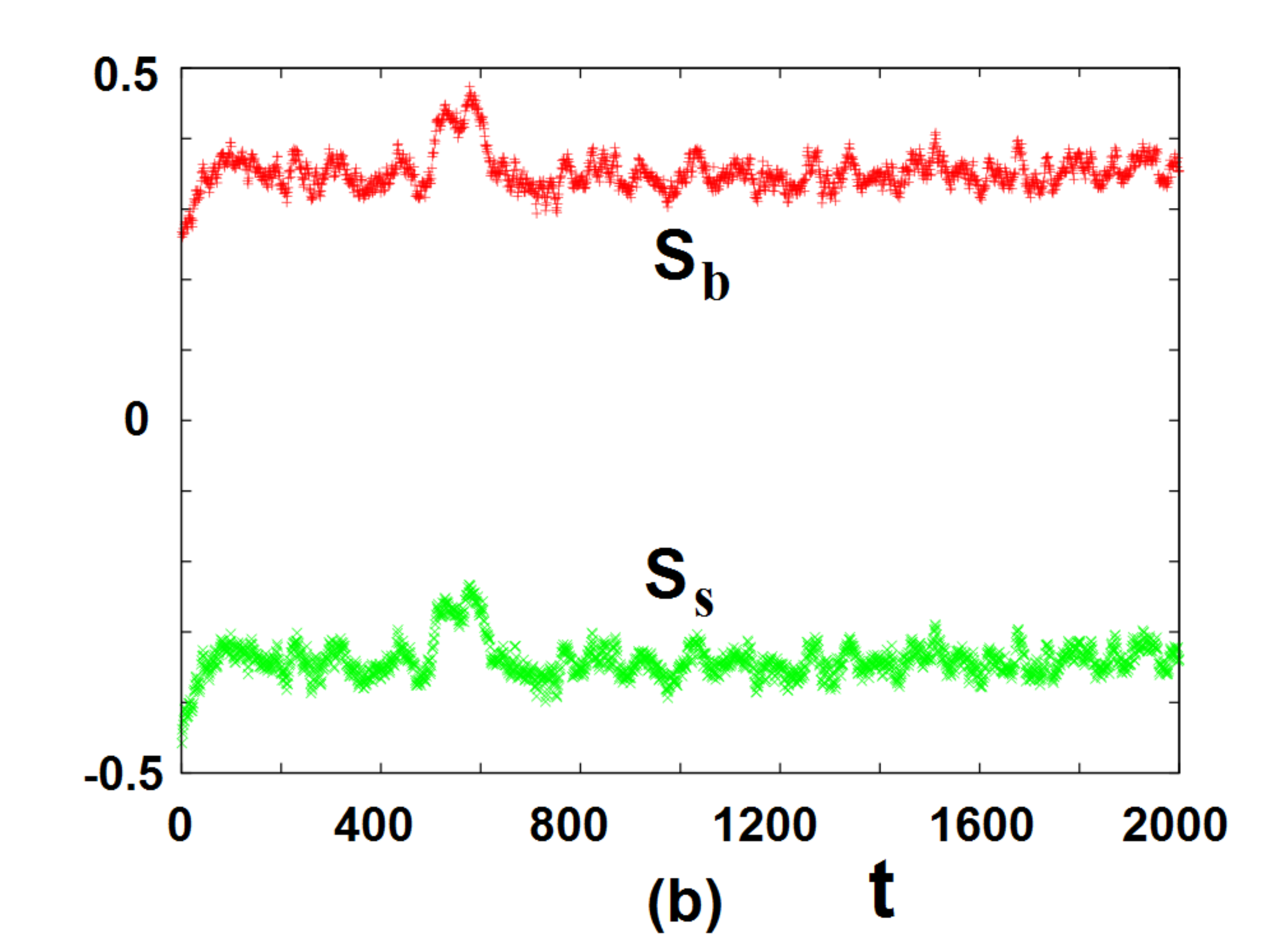}
\vspace{5pt}
\caption{Initial condition 60\% of agents who sell, 40\% of
agents who buy: effect of $H$ on the time evolution of buyer  and seller percentages, $S_b$ and $S_s$, under the boosting measure $H=0.2$ applied between $t_1=400$ and $t_2=600$, at  temperatures  (a) $T=6.692$ (just below $T_c$) and (b) $T=7.872$ (above $T_c$). Parameters: $J=1$, $a$=3.   See text for comments. }
\label{ffig3}
\end{figure}

\noindent It is interesting to examine the effect of the boosting measure
$H$.  Taking the same system as that used in Fig. \ref{ffig2}, we apply $H=0.2$  at the time $t_1=400$ and remove it at $t_2=600$. A striking effect is observed in Fig. \ref{ffig3}:\\

-if $T$ is just below $T_c$, namely $T$ is in the critical region, the boosting effect lasts for a long time after $t_2$ (see Fig. \ref{ffig3}a).\\

-if $T$ is far below $T_c$ or well above $T_c$ (Fig. \ref{ffig3}b) the boosting measure goes away as soon as it is removed.\\

\begin{figure}[h!]
\vspace{5pt}
\centering
\includegraphics[scale=0.20,angle=0]{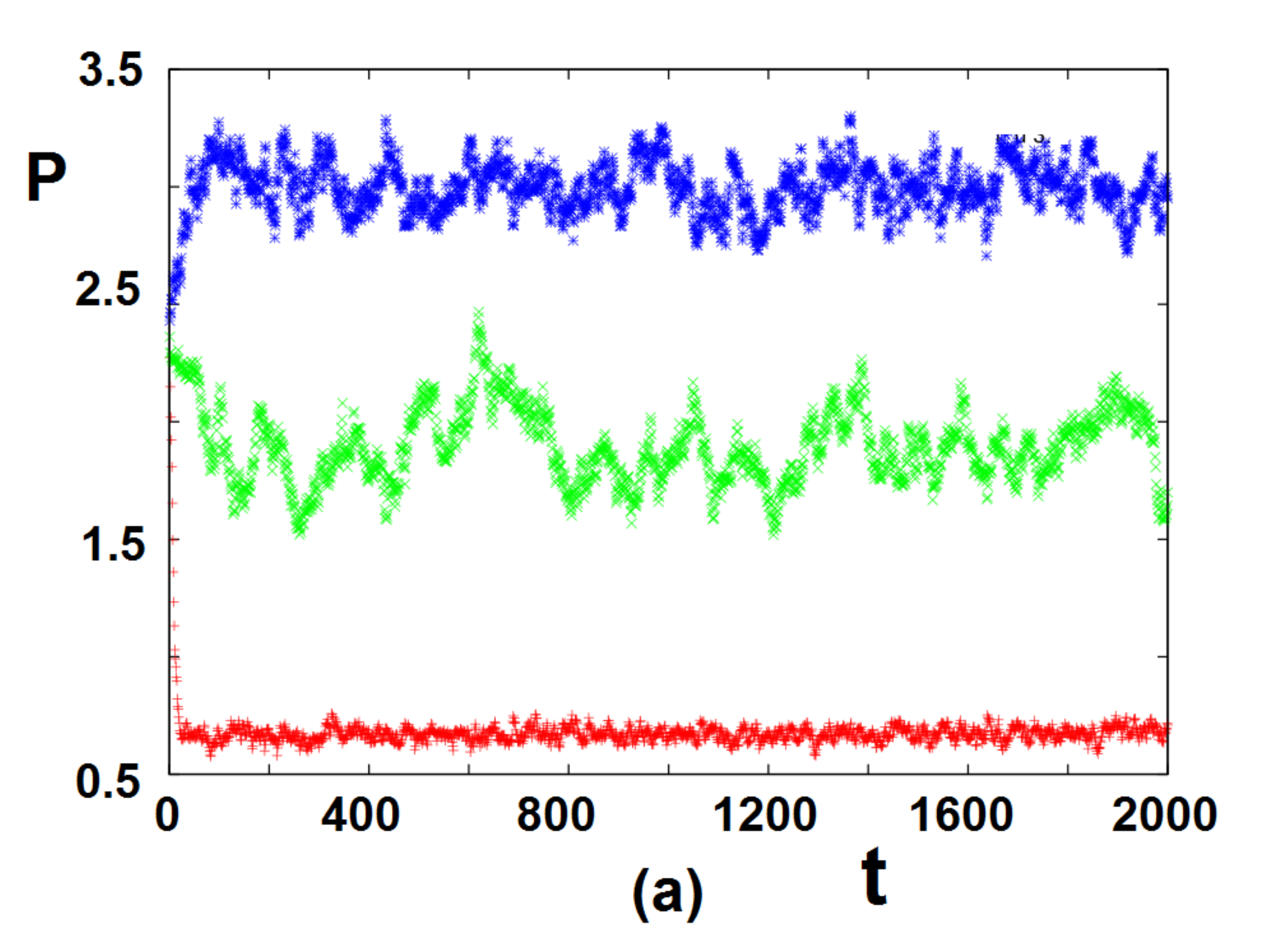}
\includegraphics[scale=0.20,angle=0]{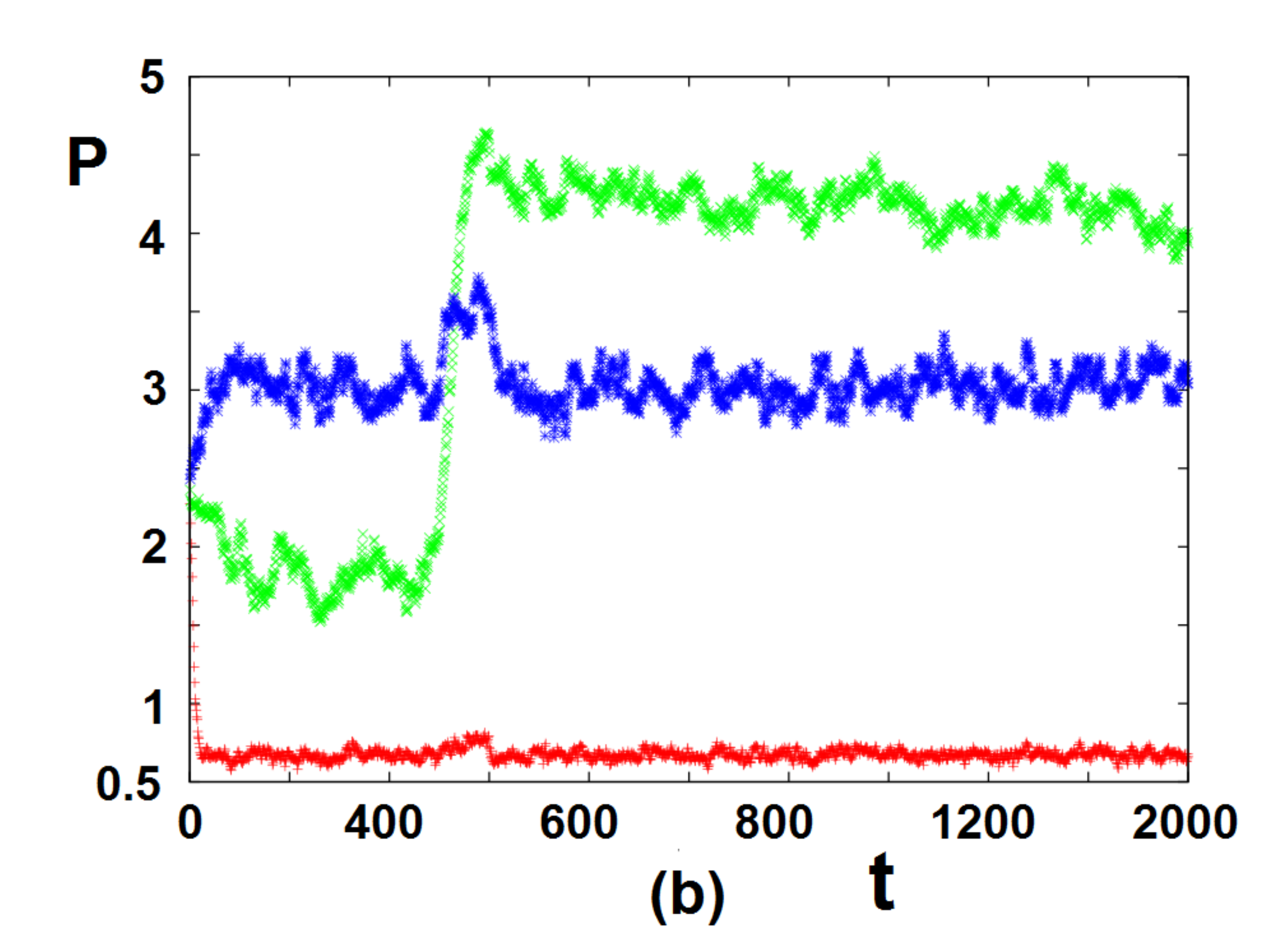}
\vspace{5pt}
\caption{Time evolution of price $P$ with initial condition 60\%  sellers, 40\% buyers. (a)  Price variation when $H=0$: red data are for $T=5.513$, green for $T=6.692$ (just below $T_c$), blue for $T=7.872$ (after market clearing) (b) Effect of $H=0.2$ applied between $t=400$ and 600. The variation of the price $P$ is shown: red data are for $T=5.513$, green for $T=6.692$, blue for $T=7.872$. Parameters: $J=1$, $a=3$, and the market clearing price $A$ is fixed  arbitrarily at 3, see Eq. (\ref{price}).  }
\label{ffig4}
\end{figure}

\noindent This effect is very interesting because although the model is simple, it suggests that boosting
  measures from the government or an economic organization are efficient and long lasting only in the economic turbulence zone just below $T_c$. We show in Fig. \ref{ffig4} the variation of the price at three typical temperatures: well below $T_c$, near $T_c$ and above $T_c$. At low $T$ the $H$ effect is not significant, at high $T$ the effect disappears when $H$ is removed. Near $T_c$, the boosting effect remains unchanged after the removal of $H$.\\

\noindent The time evolution of the price $P$ is shown in Fig. \ref{ffig4} at several market temperatures with and without a boosting measure.  These curves result from the percentage variations of buyers and sellers shown in Fig. \ref{ffig3}. A discussion was given there.\\

\begin{figure}[h!]
\vspace{5pt}
\centering
\includegraphics[scale=0.20,angle=0]{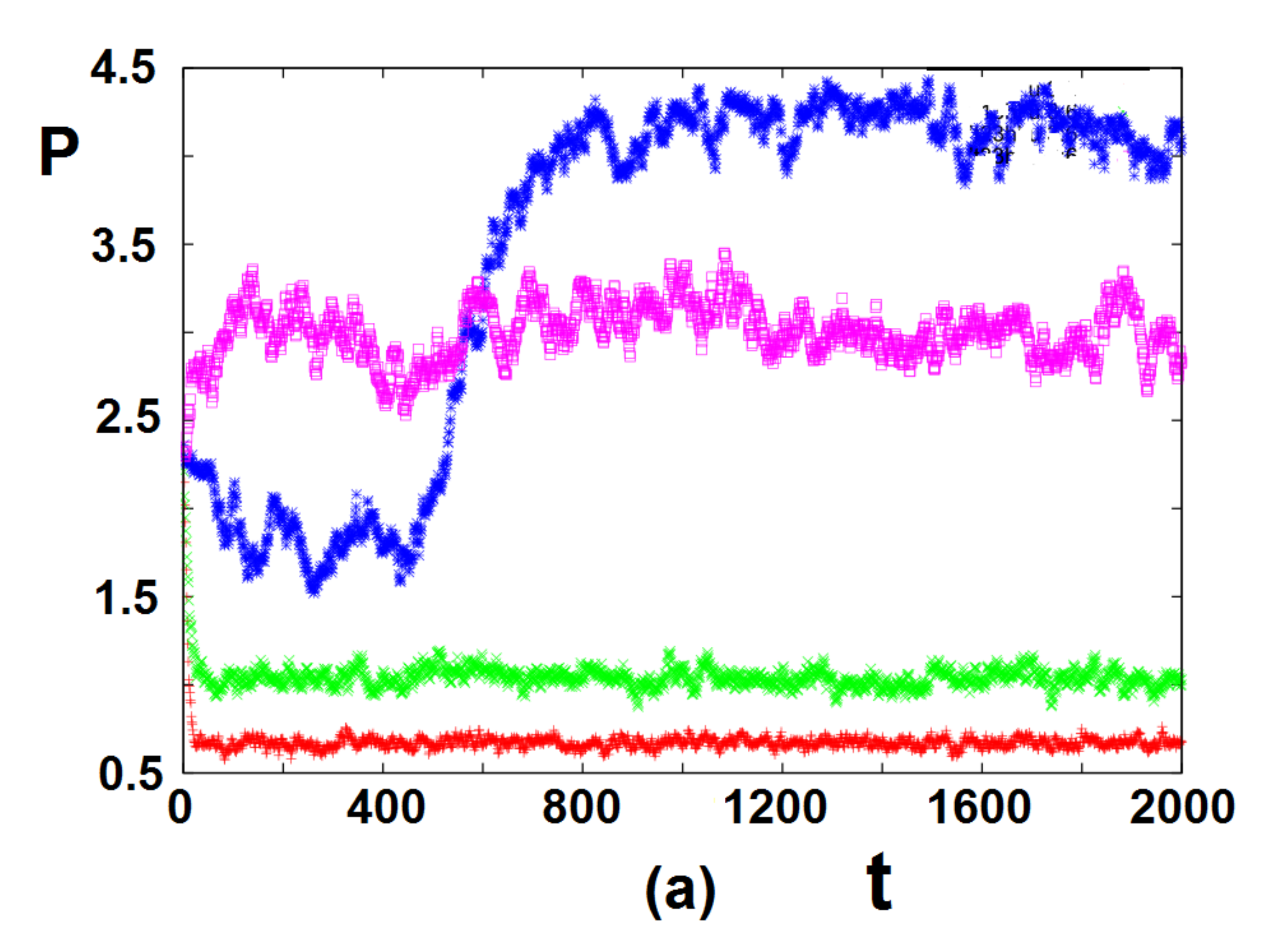}
\includegraphics[scale=0.20,angle=0]{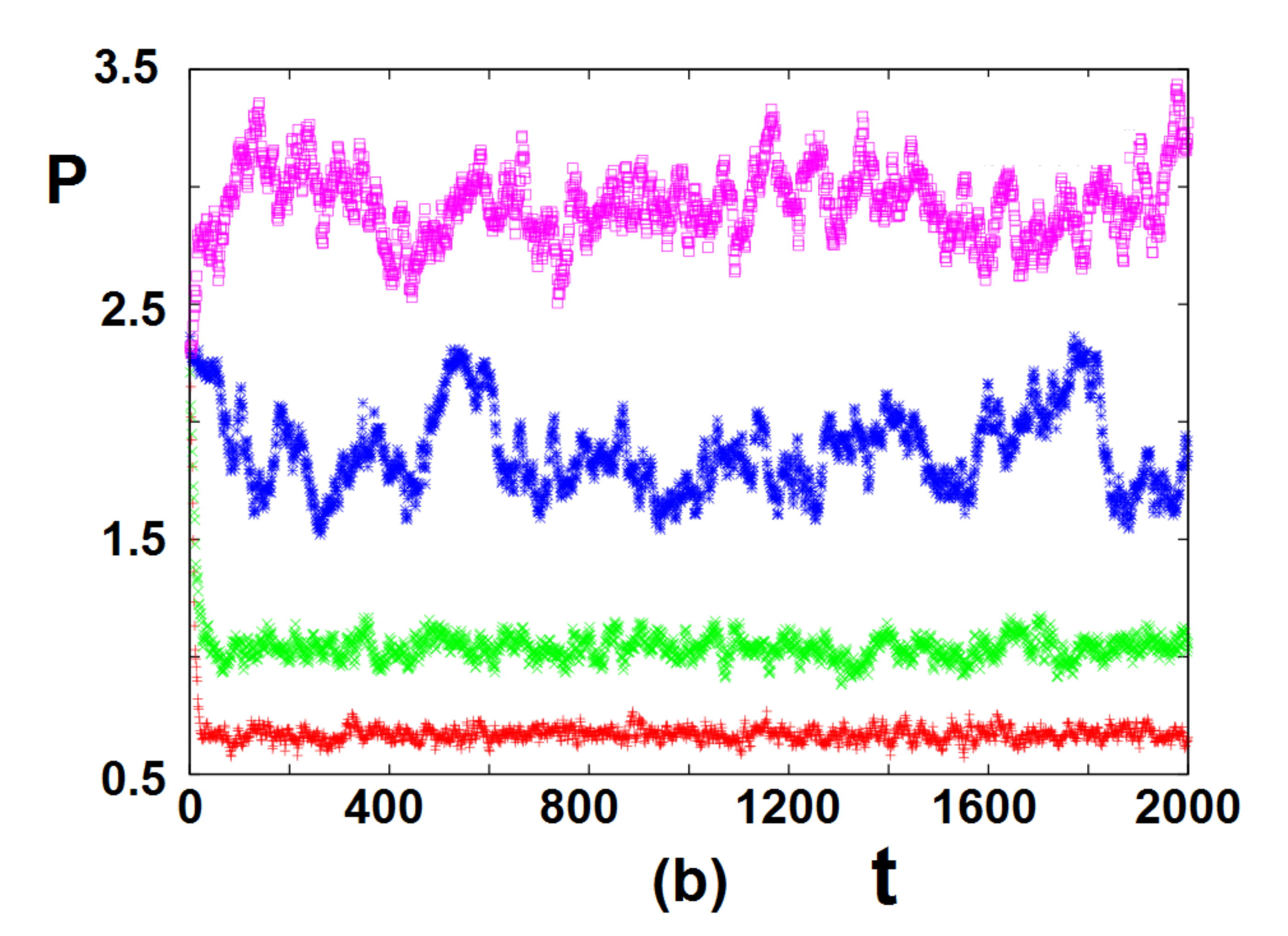}
\vspace{5pt}
\caption{Time evolution of price $P$ with initial condition 60\% of agents who sell, 40\% of
agents who buy. Red, green, blue and magenta symbols are for $T=6.103$ (well below $T_c$), $T=6.692$ (just below $T_c$),  $T=7.282$ and $T=7.872$ (after the market clearing) (a) $H=0.05$; (b) $H=0.02$.  See text for comments. }
\label{ffig5}
\end{figure}

\noindent At this stage it is interesting to examine the effect of the amplitude of $H$. We have seen in Fig. \ref{ffig4}b the jump in the price at $H=0.2$. We show in Fig. \ref{ffig5}a the case when $H=0.05$. We see here that the long lasting jump remains. However, when $H=0.02$ (Fig. \ref{ffig5}b), that jump disappears.  It means that there is a critical value $H_c$ (between 0.02 and 0.05) above which the long-lasting jump sets in.\\

\noindent The above result is a very important finding: to boost the price for a long lasting,  the measure has to be strong enough and taken in the critical region of the market temperature (crisis, unrest) in a short lapse of time.\\

\noindent In the results shown above, we have applied $H$ in a lapse of time between $t_1=400$ and $t_2=600$. If $H$ is applied during a longer period the persistence effect does not change, as expected, after we remove $H$. However,  if $H$ is applied during a too short period of time, shorter than the relaxation time induced by $H$, the price increase is smaller and the persistence decays with time (there is no long-lasting effect shown in Figs. \ref{ffig3}a, \ref{ffig4}b and \ref{ffig5}a). This relaxation time is of the order of a dozen Monte Carlo steps.

\subsection{Case of more people who buy}
\begin{figure}[h!]
\vspace{5pt}
\centering
\includegraphics[scale=0.20,angle=0]{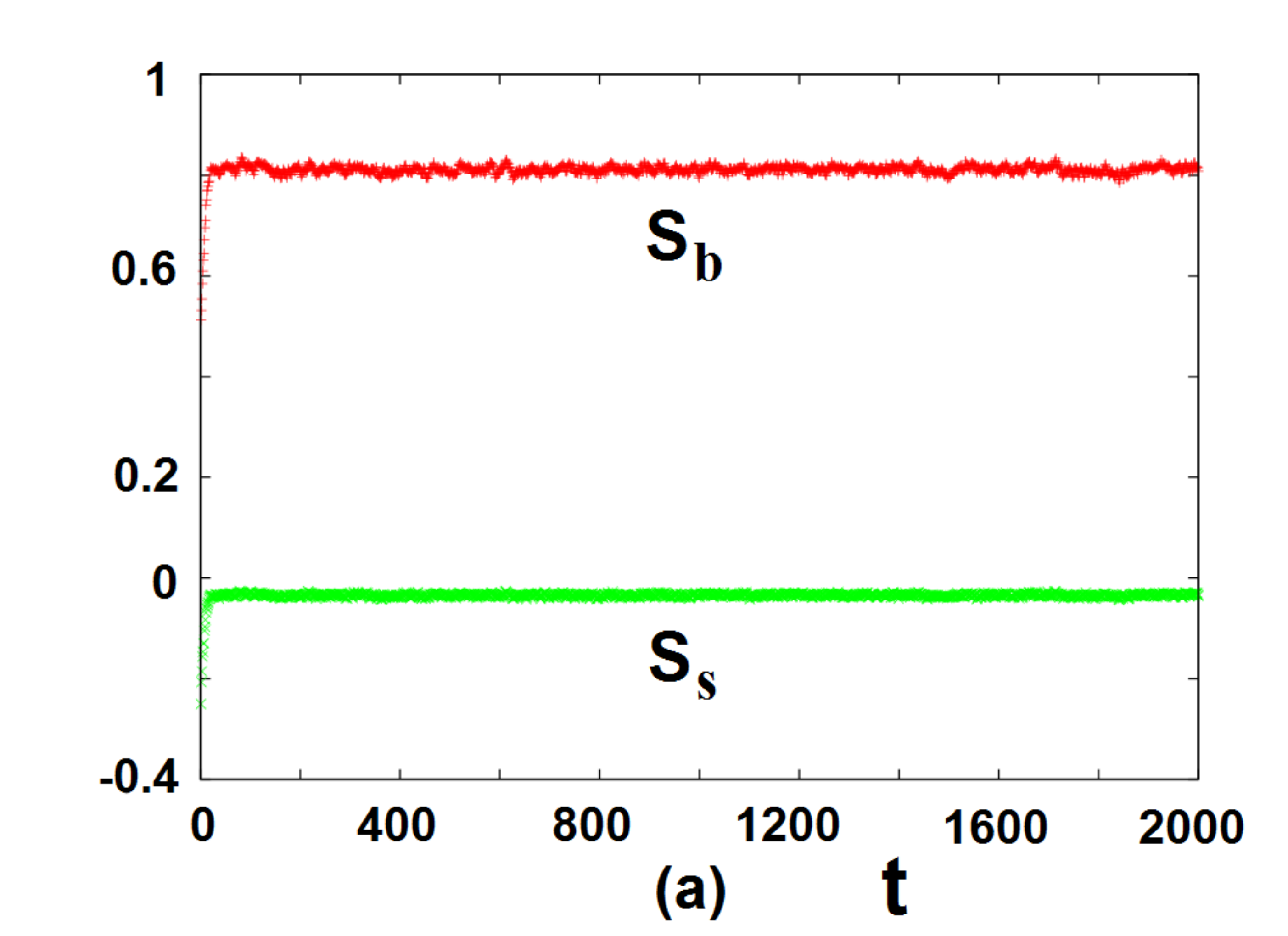}
\includegraphics[scale=0.20,angle=0]{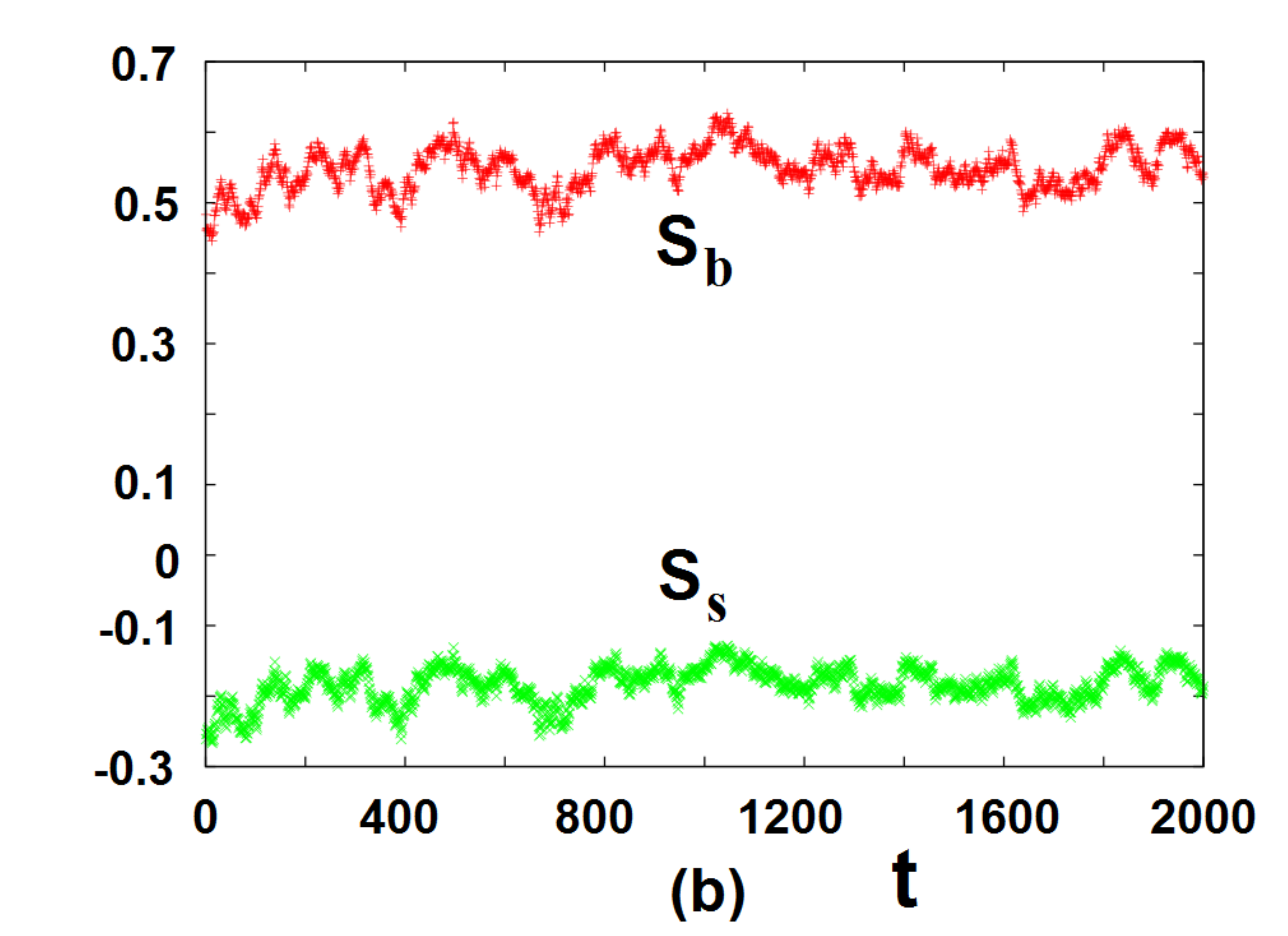}
\vspace{5pt}
\caption{Time evolution of buyer and seller percentages, $S_b$ and $S_s$, with $H=0$ and initial condition: 60\% of agents who buy, 40\% of
agents who sell. Red symbols are buyers, green symbols are sellers  (a) $T=5.513$ (well below $T_c$)(b) $T=6.692$ (just below $T_c$).  See text for comments. }
\label{ffig6}
\end{figure}

\noindent Let us take the case of more people who want to buy than people who want to sell. We start the simulation with for example 60\% buyers and 40\% sellers among the agents.
In this case, at low $T$ each agent  imitates the majority of his neighbors, leading to more people to buy. As a consequence,  the price increases, as seen  from Eq. (\ref{price}). We   show in Fig. \ref{ffig6} the time evolution of the percentages of buyers and sellers, $S_b$ and $S_s$ respectively, at several market temperatures.  One observes strong fluctuations in the critical region.\\

\begin{figure}[h!]
\vspace{5pt}
\centering
\includegraphics[scale=0.20,angle=0]{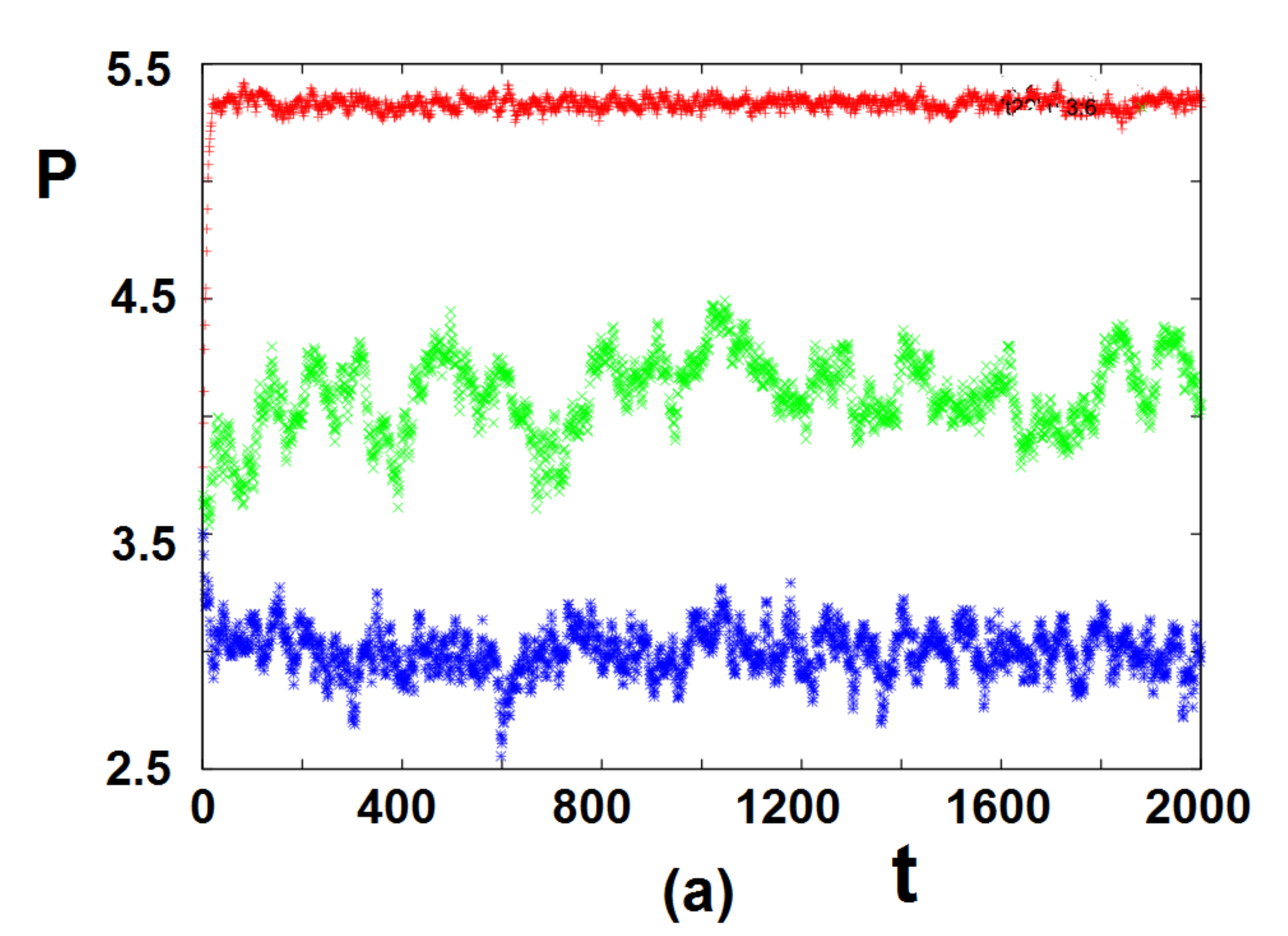}
\includegraphics[scale=0.20,angle=0]{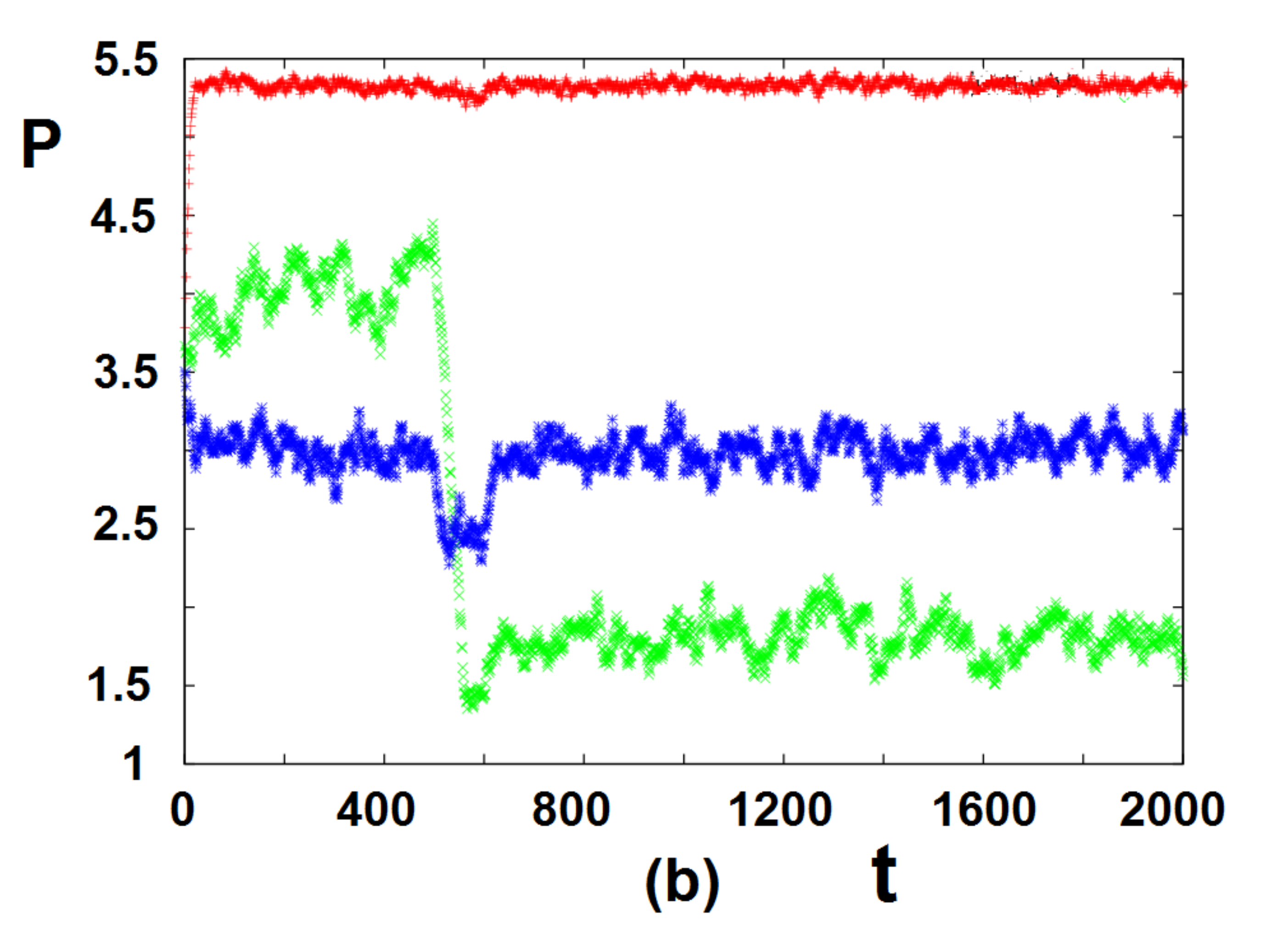}
\vspace{5pt}
\caption{Time evolution of price with initial condition 60\% of agents who buy, 40\% of
agents who sell. Red, green and blue symbols are for $T=5.513$ (well below $T_c$), $T=6.692$ (just below $T_c$) and $T=7.872$ (a) $H=0$; (b) Effect of  $H=-0.2$.  The market clearing price is fixed at $A=3$. See text for comments. }
\label{ffig7}
\end{figure}

\noindent The price $P$ as a function of time is shown in Fig. \ref{ffig7} for three typical temperatures below, near and above $T_c$. Figure \ref{ffig7}a shows the price when $H=0$. One notes that  for $T <  T_c$ the price is higher than the market clearing price fixed at 3.  Figure \ref{ffig7}b shows the effect of $H$ applied to reduce the price (with a negative value $H=-0.2$).  As seen, the effect of $H$ is not significant at $T \ll  T_c$ and $T > T_c$. However just below $T_c$, one observes a striking effect of $H$: a strong fall of price and this fall remains unchanged after the removal of $H$.  This feature has been found above for the case of more people to sell than people to buy: we found a long-lasting price jump just below $T_c$.  As the previous case, $H$ should be stronger than a critical value for a long-lasting jump to occur.

\subsection{Model of 5 individual states}

\begin{figure}[h!]
\vspace{5pt}
\centering
\includegraphics[scale=0.05,angle=0]{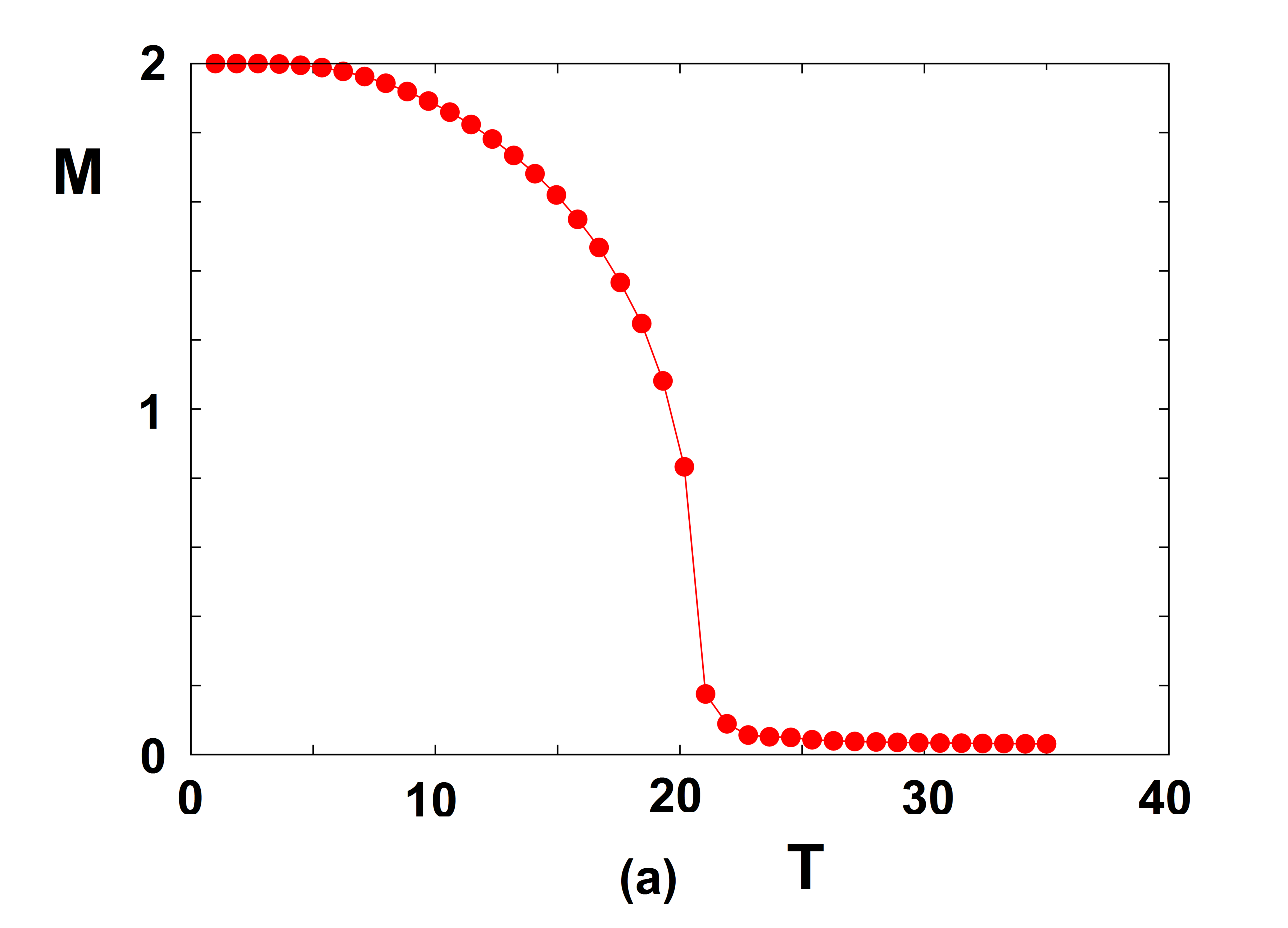}
\includegraphics[scale=0.05,angle=0]{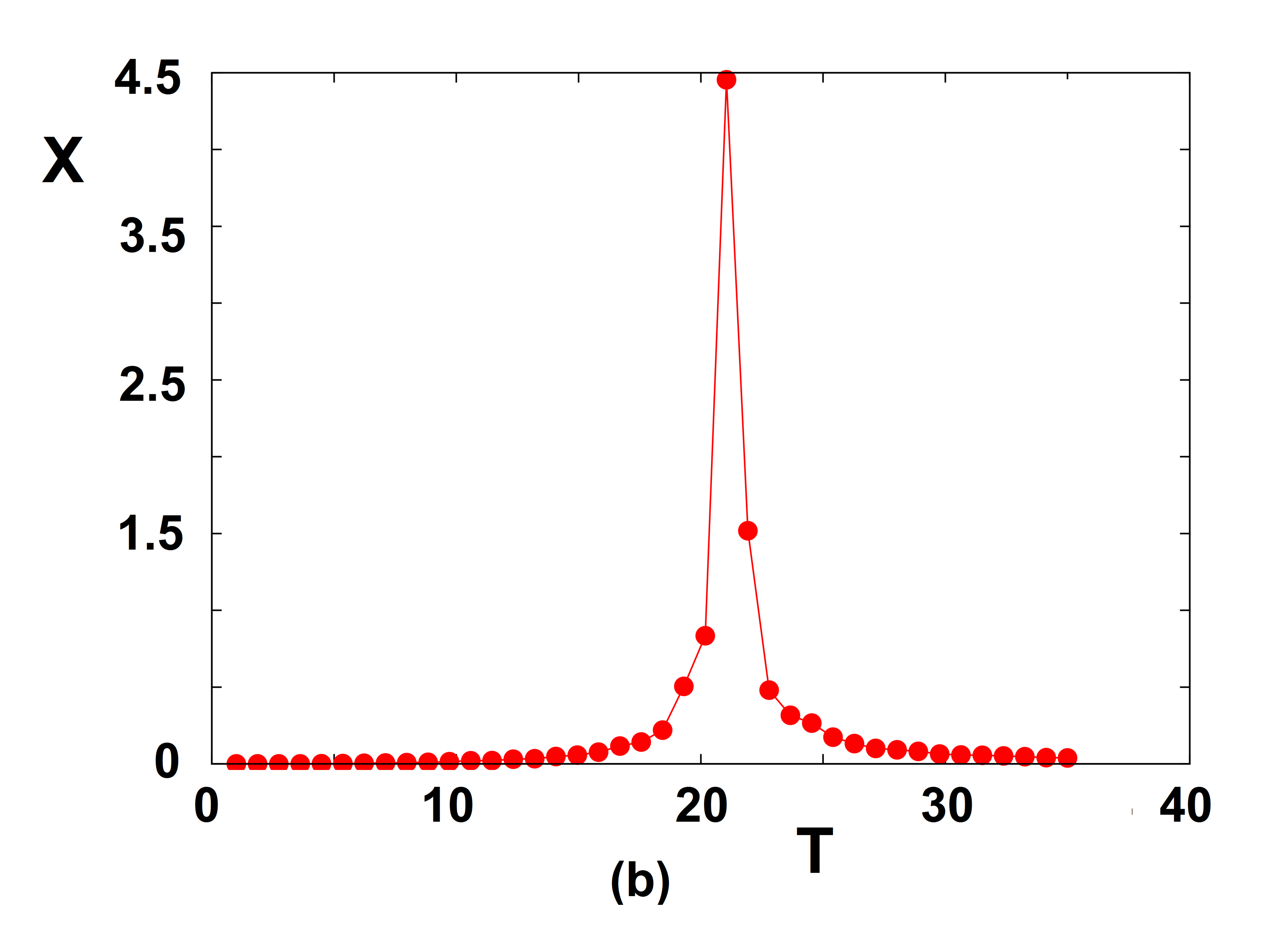}
\vspace{5pt}
\caption{Model of 5 individual states (a) Order parameter $M$, (b) susceptibility, versus $T$.  Parameters in Eq. (\ref{energy}): $J=1$, $a$=6, $H=0$.  See text for comments. }
\label{ffig8}
\end{figure}

\noindent Let us consider the 5-state model: -2, -1, 0, 1, 2. Negative values express two degrees of selling desire (strong and moderate), while positive values express two degrees of buying desire.  The market-temperature effect is shown in Fig. \ref{ffig8} where one recognizes the critical value $T_c=21.051$ for parameters $J=1$, $a$=6, $H=0$. Note that the susceptibility $\chi$ expresses the fluctuations of the order parameter $M$.\\

\begin{figure}[h!]
\vspace{5pt}
\centering
\includegraphics[scale=0.05,angle=0]{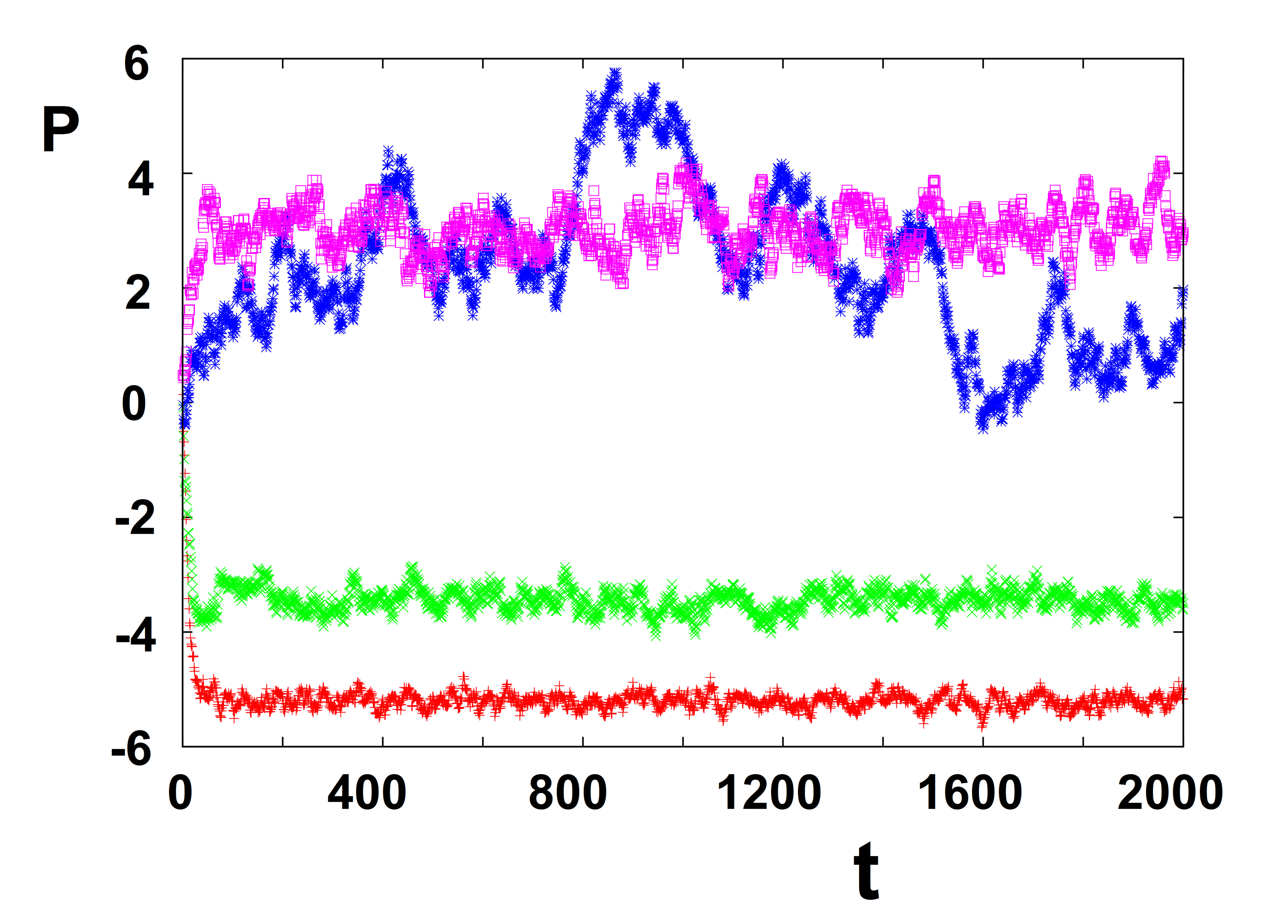}
\vspace{5pt}
\caption{
Price versus time at four temperatures $T_1=17.764 <  T_c$ (red), $T=19.308 <  T_c$ (green), $T=21.051=T_c$ (blue), $T_4=22.795 > T_c$ (magenta).  See text for comments. }
\label{ffig9}
\end{figure}

\noindent The case of less people to buy (40\%) than people to sell is initially started, namely the initial price is lower than the market clearing price. An example of the price fluctuations is shown at several $T$ in Fig. \ref{ffig9}: At low $T$, agents follow mainly the majority of their neighbors' desire so that the price fluctuations are small  (red and green curves). At $T_c$ there are strong price fluctuations (blue curve) and at $T > T_c$, the price is stabilized at the market clearing, equal to 3 in this example (magenta curve). These features are common with the 3-state case seen above.\\

\begin{figure}[h!]
\vspace{5pt}
\centering
\includegraphics[scale=0.05,angle=0]{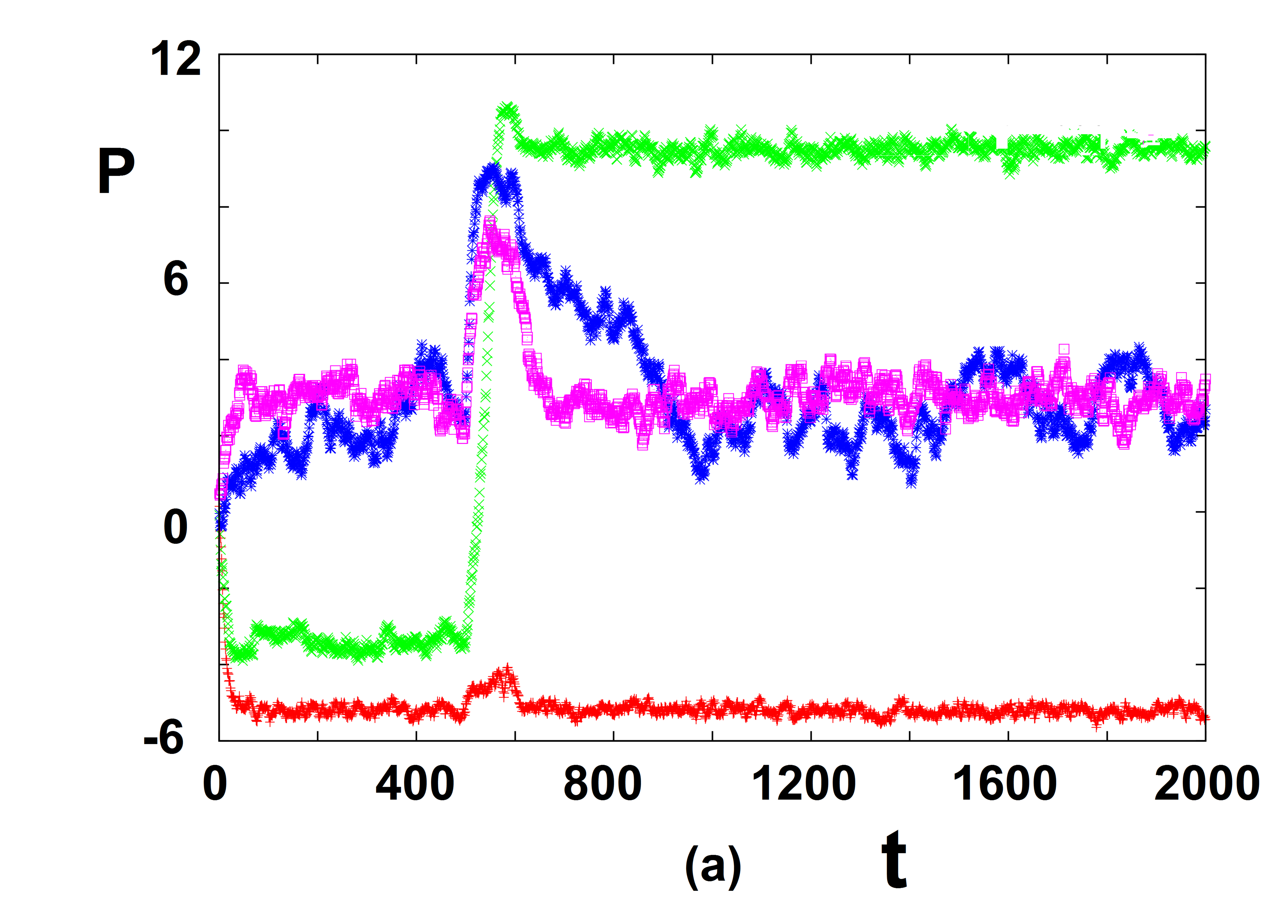}
\includegraphics[scale=0.05,angle=0]{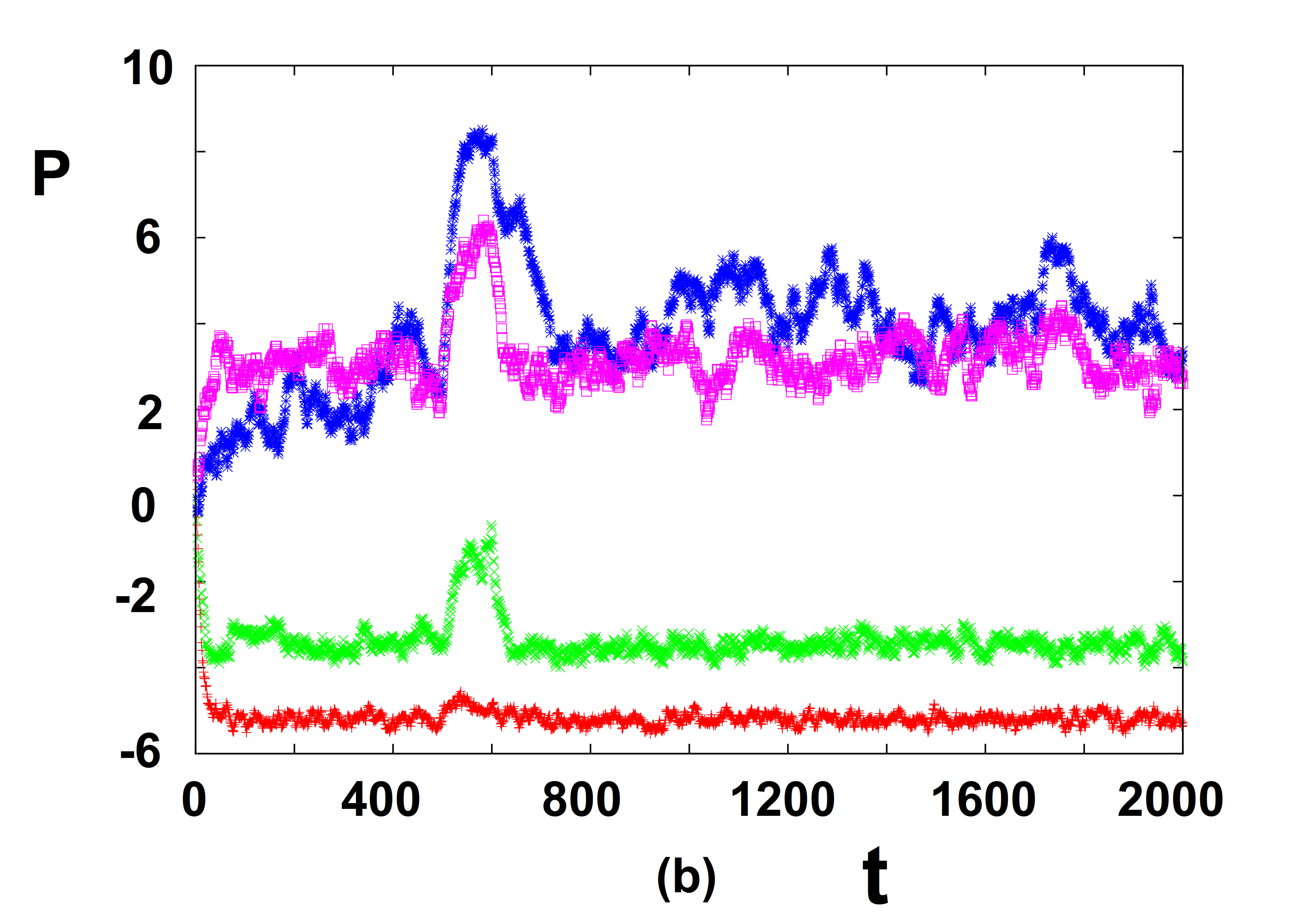}
\vspace{5pt}
\caption{Effect of the boosting measure $H$: time evolution of price with initial condition 60\% of agents who sell and 40\% of agents who buy, for  (a) $H=0.6$; (b) $H=0.35$. Color code: red for $T=17.564$, green for $T=19.308$ (just below $T_c$, blue is for $T=21.051$, magenta is for $T=22.795$. Note that $T_c=21.144$. See text for comments. }
\label{ffig10}
\end{figure}

\noindent We examine now the effect of boosting measure $H$. We show in Fig. \ref{ffig10} the time evolution of the price for several values of $H$.  Fig. \ref{ffig10}a shows the case of a strong value of $H$, namely $H=0.6$, at several $T$. As seen, the boosting measure lasts only during its application except when $T$ is near $T_c$ where the effect seems to last forever (green curve).  When we decrease $H$ down to $H=0.4$, the same effect is still observed (not shown). At $H=0.35$ the boosting falls down when $H$ is removed (Fig. \ref{ffig10}b).  Thus, there is a critical value between $H=0.35$ and 0.4 above which the boosting effect is long-lasting.\\

\noindent The same feature is found when we start with more buyers, namely the initial price at $T <  T_c$ is higher than  the market clearing price. The specific measure applied to reduce the price has a long-lasting effect only in the region just below $T_c$ if $|H|$ is larger than a critical value between 0.35 and 0.40.

\subsection{Continuous model}\label{cont}
\noindent Let us consider the case where the each agent has a continuous spectrum of states going from a strong selling desire $S=-1$ to a strong buying desire $S=1$ in a continuous manner: $S\in[-1,1]$.  The determination of $T_c$ by examining the order parameter $M$ and the susceptibility $\chi$ gives $T_c=3.513$.\\

\begin{figure}[h!]
\vspace{5pt}
\centering
\includegraphics[scale=0.05,angle=0]{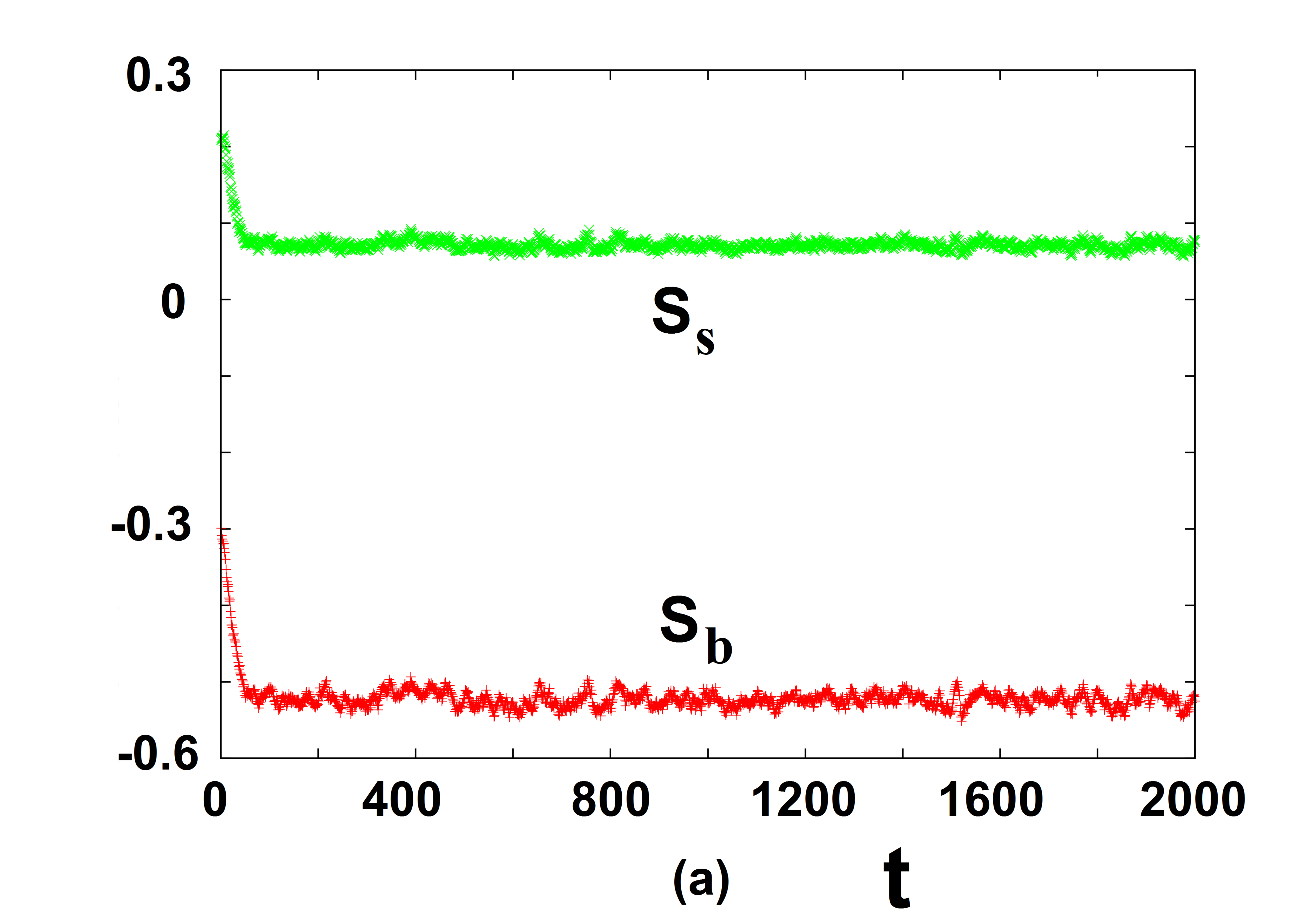}
\includegraphics[scale=0.05,angle=0]{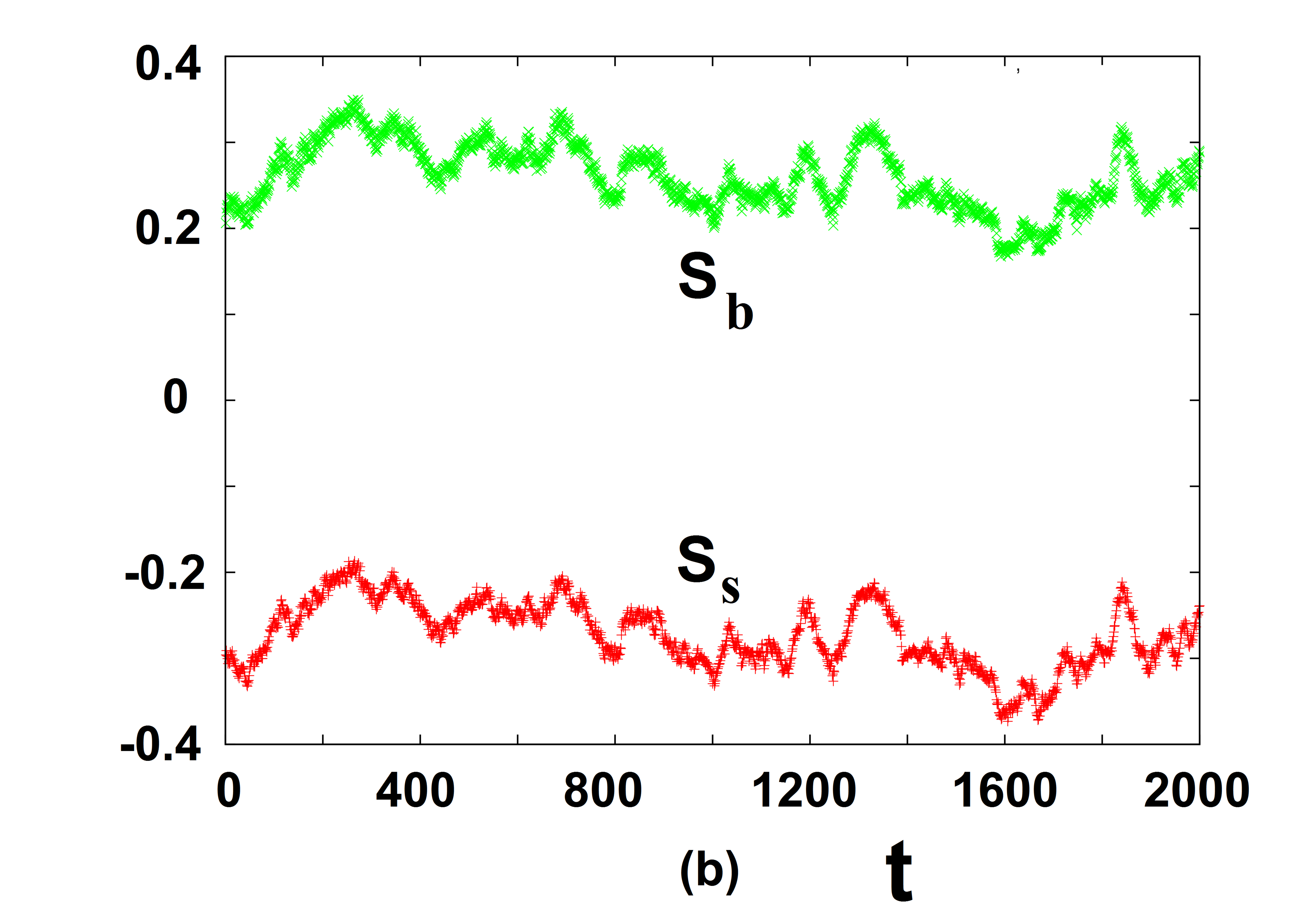}
\includegraphics[scale=0.05,angle=0]{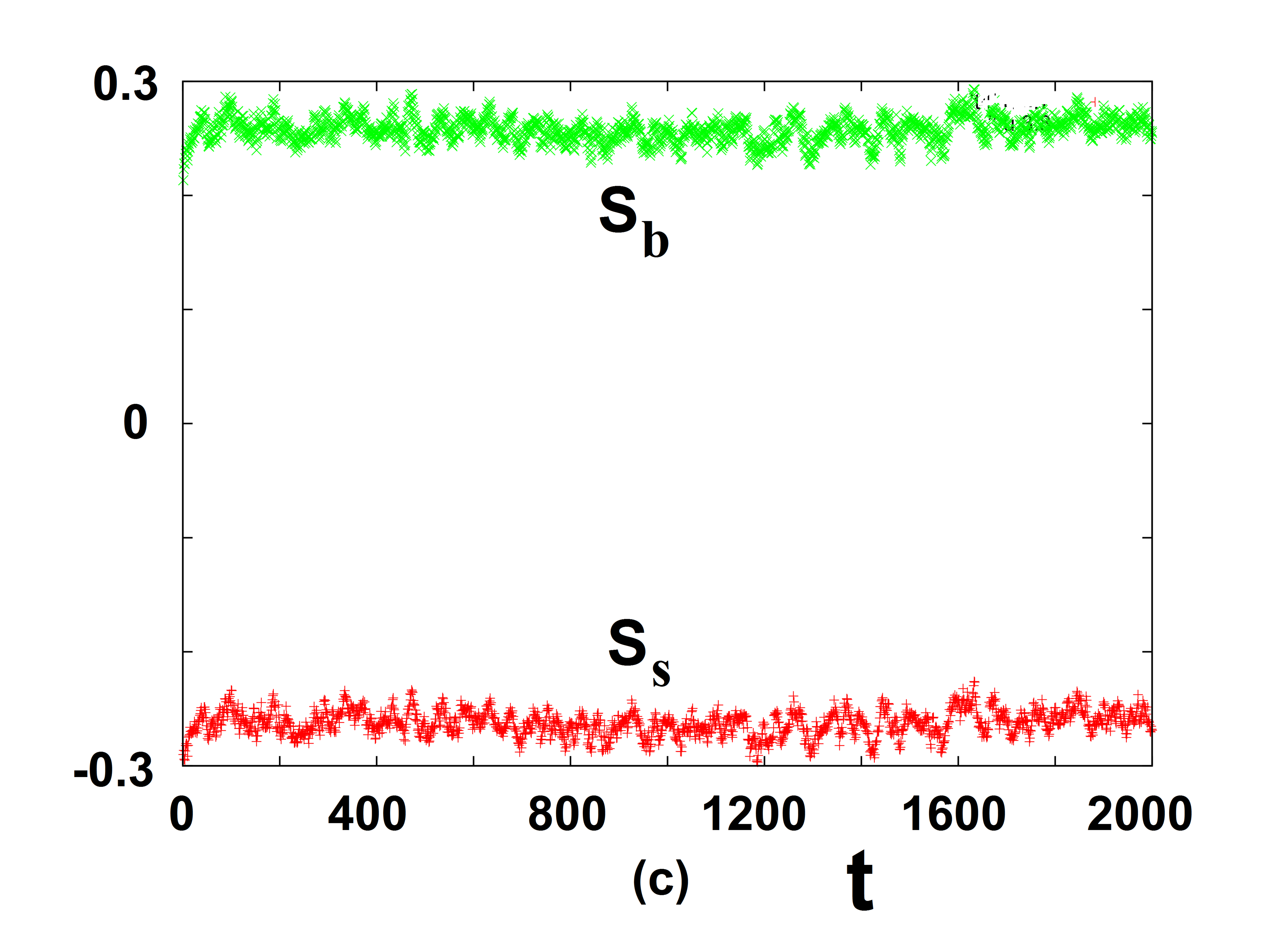}
\includegraphics[scale=0.065,angle=0]{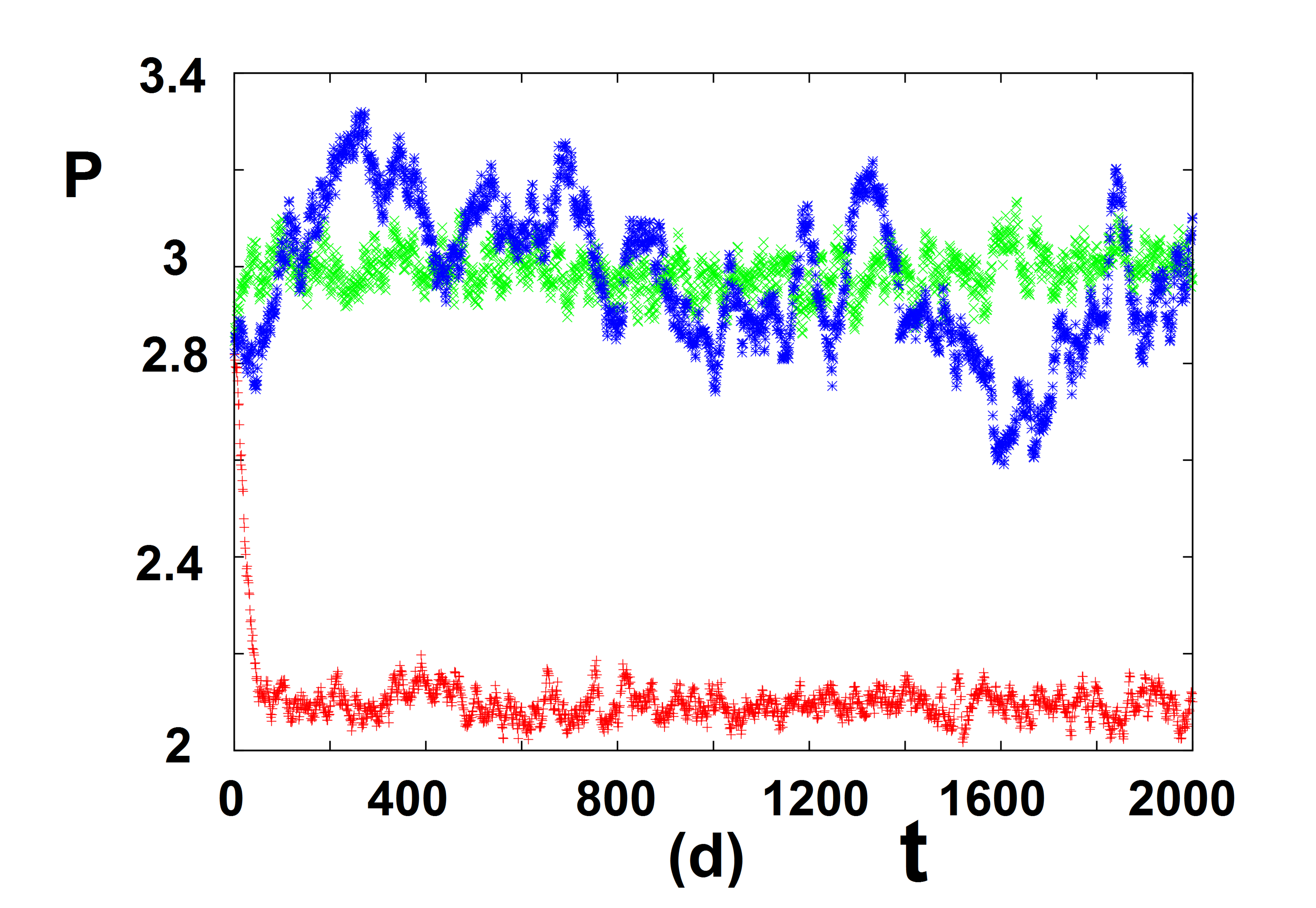}
\vspace{5pt}
\caption{(a)-(b)-(c):Time evolution of seller (red) and buyer (green) percentages, $S_s$ and $S_b$ respectively, with $H=0$ using the initial condition 60\% of agents who sell, 40\% of
agents who buy at (a) $T_1=3.256 \ll  T_c$, (b) $T_2=3.564$ just above $ T_c$, (c) $T_3=3.974 >>T_c$.
(d) Price $P$ versus time $t$ at temperatures $T_1$ (red), $T_2$ just above $T_c$ (blue), $T_3$ (green).  Market clearing price is fixed at $A=3$. See text for comments. }
\label{ffig11}
\end{figure}

\noindent We show in Fig. \ref{ffig11} the case of more sellers than buyers initially. As in the previous cases, the price is low for $T <  T_c$ and reaches the market clearing value for $T\geq T_c$. We have as before strong fluctuations  of the price at and near $T_c$(see Fig. \ref{ffig11}d).\\

\begin{figure}[h!]
\vspace{5pt}
\centering
\includegraphics[scale=0.05,angle=0]{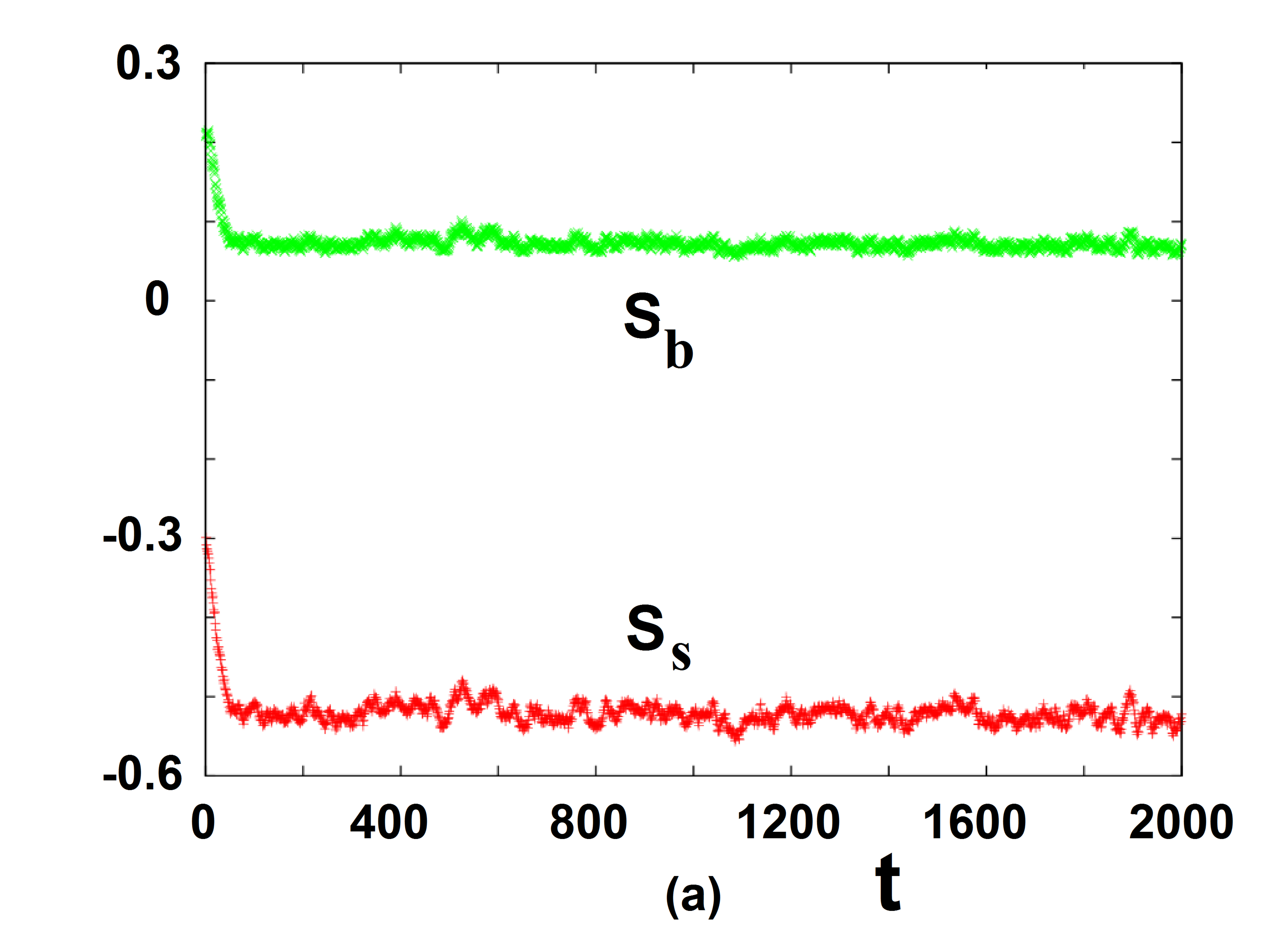}
\includegraphics[scale=0.05,angle=0]{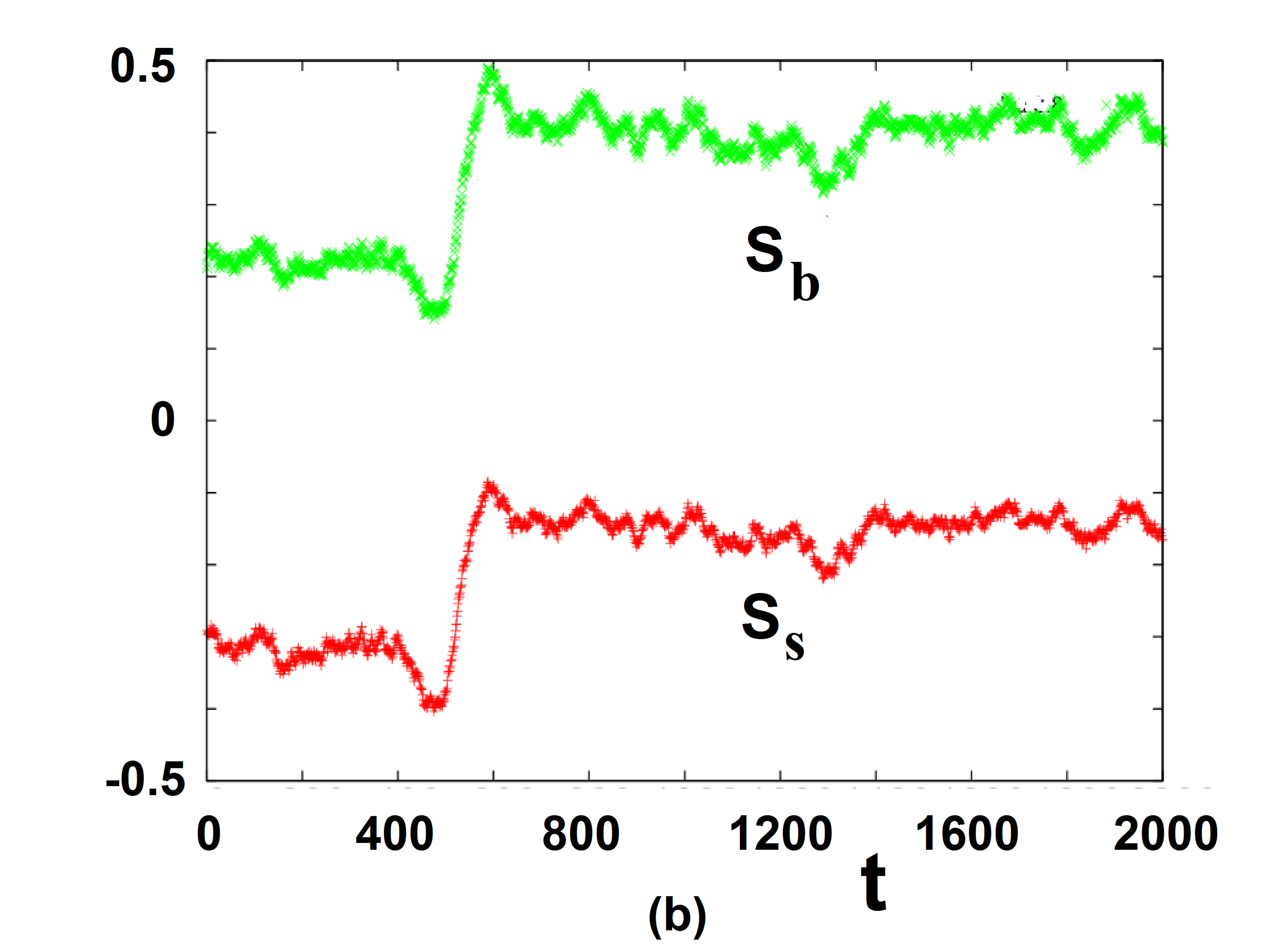}
\includegraphics[scale=0.048,angle=0]{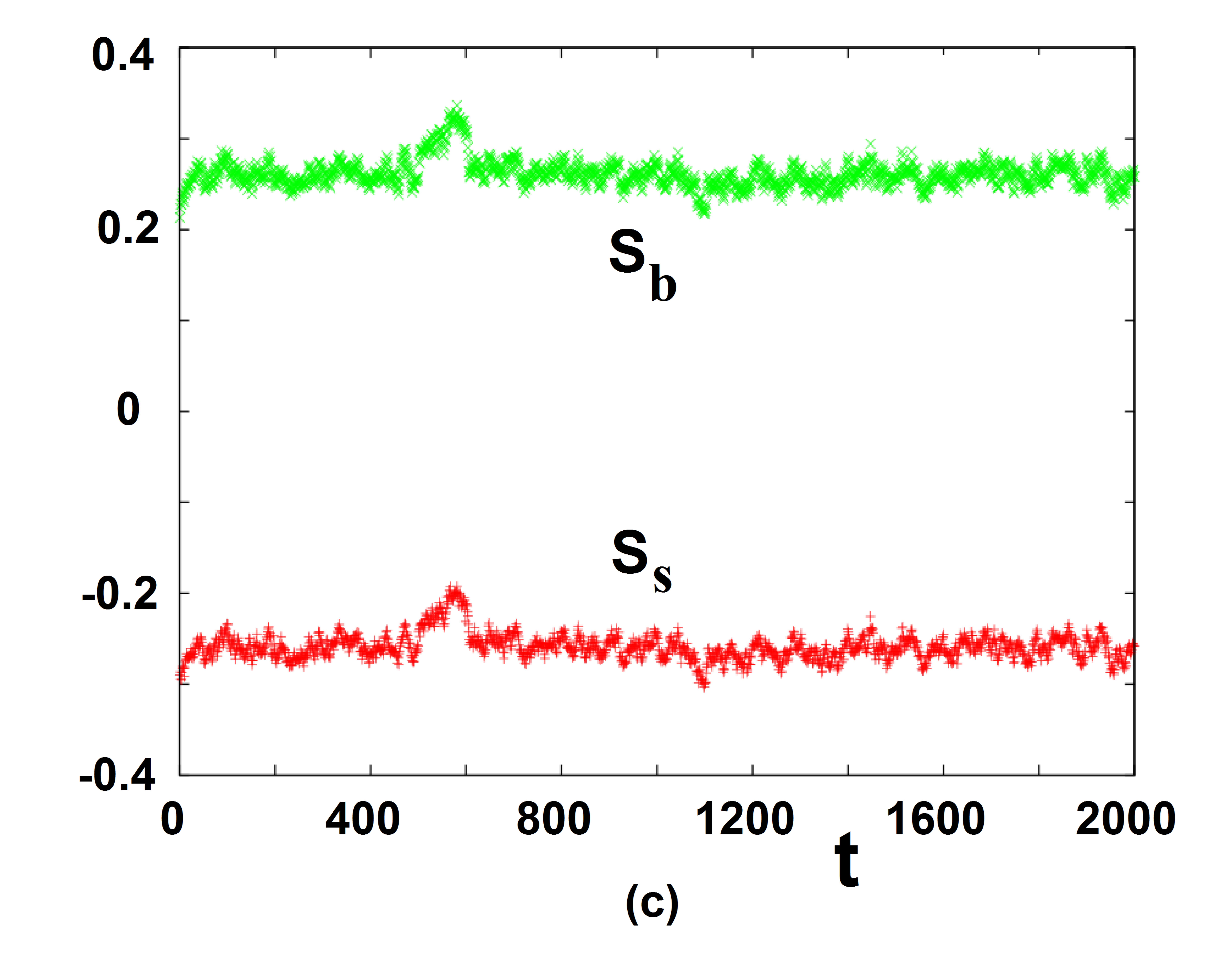}
\vspace{5pt}
\caption{Effect of the boosting measure $H$: time percentage evolution of sellers $S_s$ (red) and buyers $S_b$ (green) with initial condition  60\% of
agents who sell and 40\% of agents who buy for $H=0.1$ at (a) $T=3.256 \ll  T_c$, (b) $T=3.462$ just below $ T_c$, (c) $T=3.974  \gg T_c$.   Note the jump observed at $T=3.462$. See text for comments. }
\label{ffig12}
\end{figure}

\noindent We show in Fig. \ref{ffig12} the effect of boosting measure $H$ applied between $t_1=400$ and $t_2=600$ (in MC-step time unit). The percentages of sellers $S_s$ and buyers $S_b$ make a jump at near $T_c$. This jump lasts even after the removal of $H$. At other temperatures far from $T_c$, the $H$ effect is insignificant and barely seen in Figs. \ref{ffig12}a  and Figs. \ref{ffig12}b, during the application of $H$.\\

\begin{figure}[h!]
\vspace{5pt}
\centering
\includegraphics[scale=0.05,angle=0]{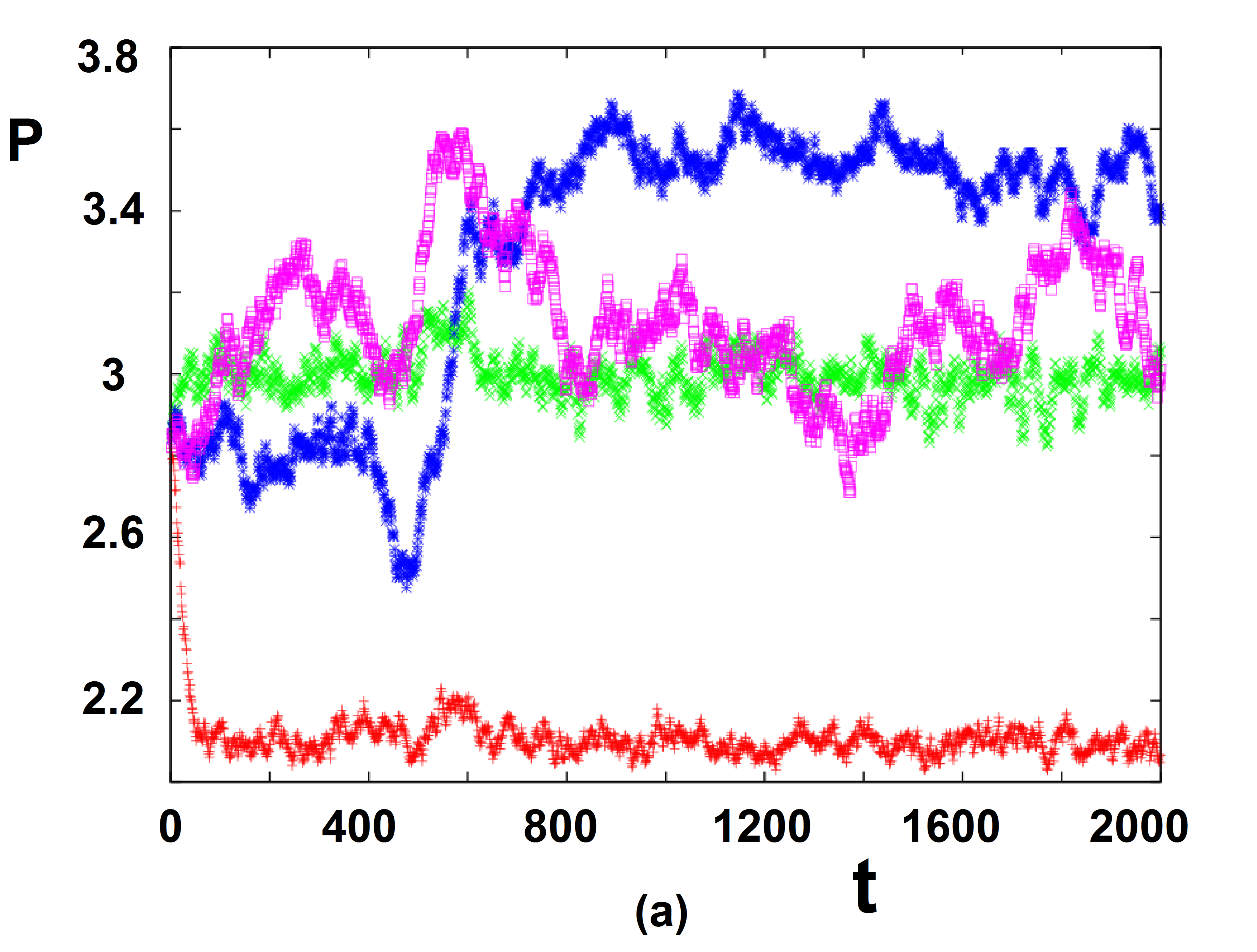}
\includegraphics[scale=0.05,angle=0]{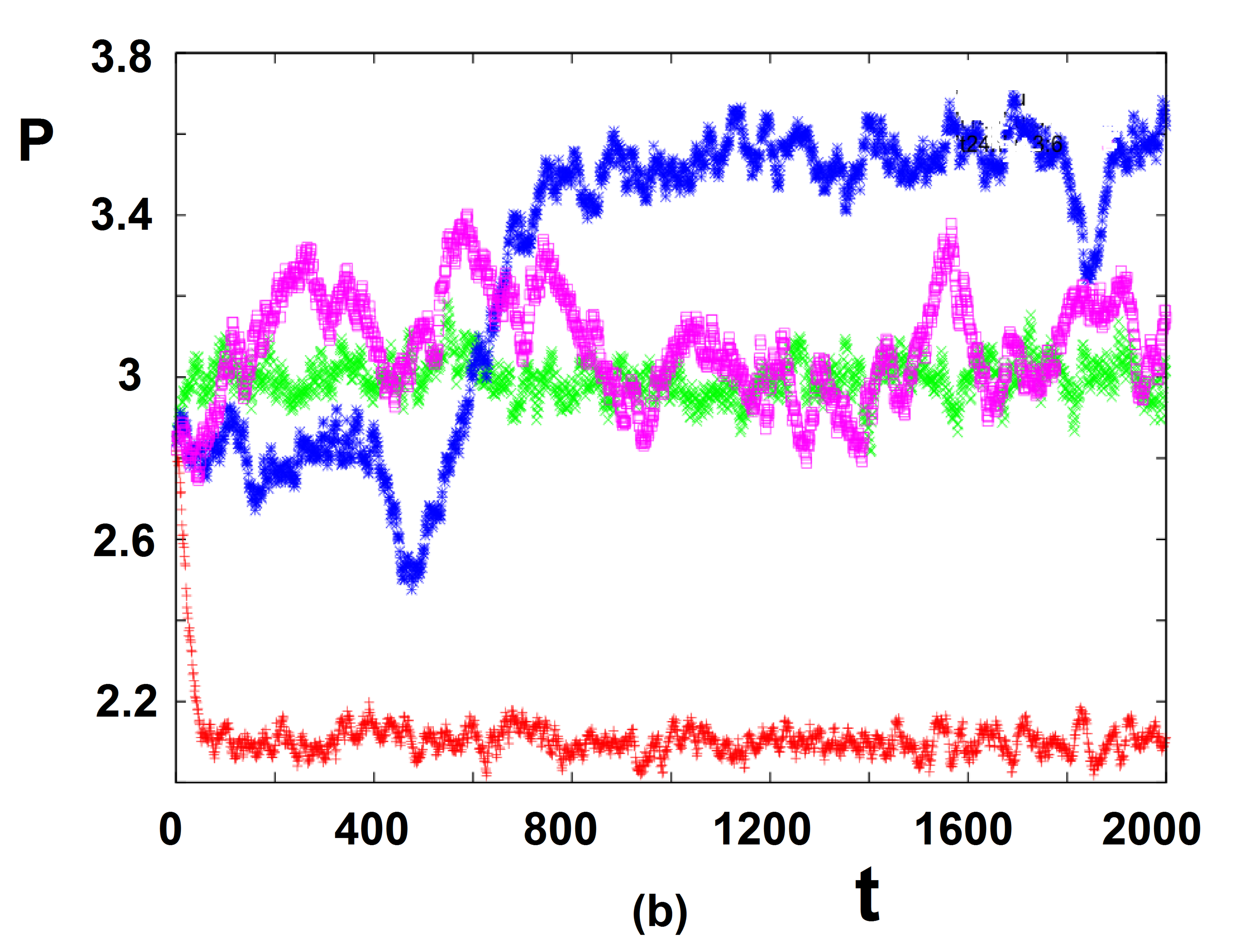}
\includegraphics[scale=0.048,angle=0]{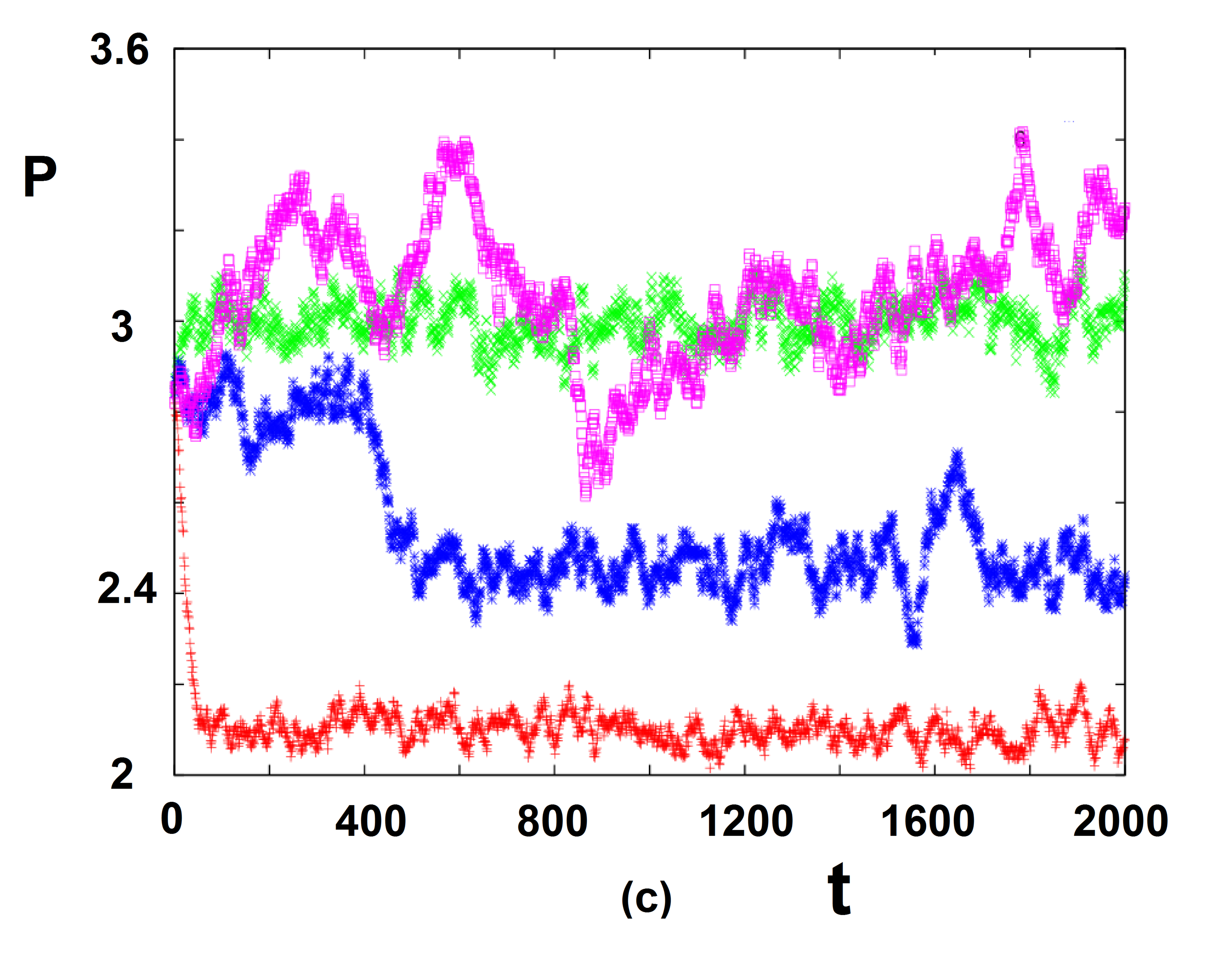}
\vspace{5pt}
\caption{Effect of the boosting measure $H$ on the price, with initial condition  60\% of
agents who sell and 40\% of agents who buy for (a) $H=0.06$, (b) $H=0.025$, (c) $H=0.02$. Color code for temperatures: red is for $T=3.256 \ll  T_c$, blue is for $T=3.462$ just below $T_c$, magenta is for $T=3.564$ just above $ T_c$, green is for $T=3.974 \gg T_c$.  See text for comments. }
\label{ffig13}
\end{figure}

\noindent The effect of $H$ on the price reflects the variations of seller and buyer percentages as shown in Fig. \ref{ffig13} for three values of $H$: 0.06, 0.025 and 0.02.  At each value of $H$ four temperatures are examined, two below and two above $T_c$. While for $H=0.06$ and 0.025 a long-lasting jump of the price is seen at $T$ just below $T_c$, the case of $H=0.02$ does not show such a jump. The critical value of $H$ is thus between 0.02 and 0.025.

\subsection{Summary and discussion}\label{MCdiscussion}

\noindent Let us summarize some main results of our model treated by Monte Carlo simulations:
\begin{itemize}
\item The results show that the main features do not depend on the number of individual states $q$: we have similar qualitative results for $q=3$, 5 and $\infty$.
\item We have shown the primordial role of the economic temperature $T$: there is a critical value $T_c$ corresponding to the market clearing point. At and above $T_c$ the two populations buyers and sellers are equal, the price is stabilized.
\item In the region just below $T_c$ the fluctuations are very strong, namely strong exchange of stock market shares. In this region, volatility is much higher. More generally, different temperatures correspond to different volatilities, and infrequent changes in the temperature generate volatility clustering (a well-known stylized fact).  The long autocorrelation occurs in this ``critical" region is well-known in statistical physics as the critical-slowing-down phenomenon occurring near the second-order phase transition \cite{DiepSP}.
\item It is also in the region below $T_c$ that under a temporary shock the price can have a persistent effect if the shock is strong enough.  In a view point from statistical physics, the persistent state is related to a metastable state: the field $H$ drives the system to a local minimum in the free-energy landscape. The system stays in that minimum for a long time because the barrier is so high to allow the system to climb up to get out.
\end{itemize}

\noindent The above results are interesting because they come from a single energy model [Eq. (\ref{energy})] without approximation.

\section{Mean-field theory}\label{MFT}

\noindent The mean-field theory (MFT) is a very popular method in statistical physics. Its principle consists in taking into account only the average values of the neighbors acting on a spin, neglecting real-time fluctuations of each neighbor. In doing so, the calculation of thermodynamic properties is simple and the main features can be easily obtained.

\subsection{Mean-Field Model}
\noindent In this section, we use a time-dependent MFT, namely a spin at the time $t$ interacts with its neighbors of the same community in their state at $t-1$ and interacts with the average value of the other community at $t-1$.   We recall that a spin $\sigma_i(t)$ describes the state of an agent $i$ at $t$.  The time-dependent MFT has been used to study social conflicts \cite{Diep2017,Kaufman3,Diep2019}. We follow the same method in the present paper.\\

\noindent In order to make our problem of price variation richer, we consider in this section a more general situation which consists of introducing the following hypotheses:

\begin{itemize}
\item The buyers and the sellers belong to two distinct communities. The intra-community interactions are different, namely $J_1$ for the buyer community and $J_2$ for the seller. Note that in Eq. (\ref{energy}), we have only $J$ for both. This two-community hypothesis expresses the fact that in reality the moral of willing to sell or to buy is different. Sellers and buyers do not share the same enthusiasm.

\item The interaction between the communities 1 and 2, namely $K_{1,2}$, does not need to be equal to $K_{2,1}$. This hypothesis means that ``you want to sell, but I do not want to buy". So in general $K_{1,2}\neq K_{1,2}$. They are different in magnitude and in sign.
\end{itemize}

\noindent For our present problem with the hypotheses described above, the energy $E_{i}$ of an agent of group $i$ ($i=1,2$) at time $t$  is written as

\begin{eqnarray}
E_1&=& s_1(t)\left \{-j_1 s_1(t-1)-k_{12}s_2(t-1)
+a\left [s_1(t-1)-s_2(t-1)\right ] \right \}\label{MF1}\\
E_2&=& s_2(t)\left \{-j_2  s_2(t-1)-k_{21}s_1(t-1)
+a\left [s_1(t-1)-s_2(t-1)\right ] \right \}\label{MF2}
\end{eqnarray}
where we introduce the lag time in the above equations by letting the preference $s_1$ and $s_2$ at time $t$ interact with the averages $s_1$ and $s_2$ evaluated at an earlier time $t-1$. Here time is measured in units of the delay time.\\

\noindent The value of an individual in each group at the time $t$ is given by the MFT equations

\begin{eqnarray}
s_1(t)&=&\frac{\sum_{s=-M_1}^{M_1}se^{\left\{s[j_1s_1(t-1)+k_{12}s_2(t-1)
-a(s_1(t-1)-s_2(t-1))]\right\}} }{\sum_{s=-M_1}^{M_1}e^{\left\{s[j_1s_1(t-1)+k_{12}s_2(t-1) -a(s_1(t-1)-s_2(t-1))]\right\}}}\label{eq1a}\\
s_2(t)&=&\frac{\sum_{s=-M_2}^{M_2}se^{\left\{s[j_2s_2(t-1)+k_{21}s_1(t-1)
-a(s_1(t-1)-s_2(t-1))]\right\} }}{\sum_{s=-M_2}^{M_2}e^{\left\{s[j_2s_2(t-1)+k_{21}s_1(t-1)
-a(s_1(t-1)-s_2(t-1))]\right\}}}\label{eq1b}
\end{eqnarray}
where $j_n=J_n/T$ and $k_{n,m} = K_{n,m}/T$ for $n$,$m$ = 1, 2. We use units such that $k_B = 1$.
The sums on the right hand sides of the above equations lead to the Brillouin function (p. 292 in Ref. \onlinecite{DiepSP}):

\begin{equation}\label{eq2}
B(x,y,z,j,k,l,M)=(M+\frac{1}{2})\coth[(M+\frac{1}{2})(jx+ky+lz)]-\frac{1}{2} \coth[\frac{1}{2}(jx+ky+lz)]
\end{equation}

\noindent Equations (\ref{eq1a})-(\ref{eq1b}) can be written as:
\begin{eqnarray}
s_1(t)&=&B(s_1(t-1),s_2(t-1),j_1,k_{12},M_1,a)\label{eq3a}	\\
s_2(t)&=&B(s_2(t-1),s_1(t-1),j_2,k_{21},M_2,a)\label{eq3b}
\end{eqnarray}
For a given set of $(J_1,J_2,K_{12},K_{2,1},a)$ at the market temperature $T$, we can solve numerically these equations iteratively as a function of $t$: taking input values (initial condition) for $s_1(t=0)$ and $s_2(t=0)$, we calculate $s_1(t=1)$ and $s_2(t=1)$. Using these values we calculate $s_1(t=2)$
and $s_2(t=2)$ and so on.   Note that as in the model for MC simulations, the MFT model has two separate communities: the initial value of buyers $\sigma_1(t=0)$ should be positive and that of $\sigma_2(t=0)$ should be negative, but in the MFT each group starts with 100\%.
The constant $a$ in the case of two-group MFT model is more important than in the MC model as will be seen below.\\

\noindent As the time evolves, $s_1$ and $s_2$ change.  We have to bear in mind that in the present MFT hypotheses, the average value $ \langle  s_i \rangle $ ($(i=1,2)$) means the percentage of the community $i$ with  the following convention: positive $\langle  s_i \rangle $ represents the percentage of people who want to buy,  negative $\langle  s_i \rangle $ represents that of people who want to sell, as we have used in MC model below Eq. (\ref{energy}).

\subsection{Mean-Field Results}
\noindent For simplicity, we choose $M_1=M_2=1$, namely $s=-1,0,1$ in Eqs. (\ref{eq1a})-(\ref{eq1b}) (three-state model).  Let us show now the results of the MFT.\\

\begin{figure}[h!]
\vspace{5pt}
\centering
\includegraphics[scale=0.05,angle=0]{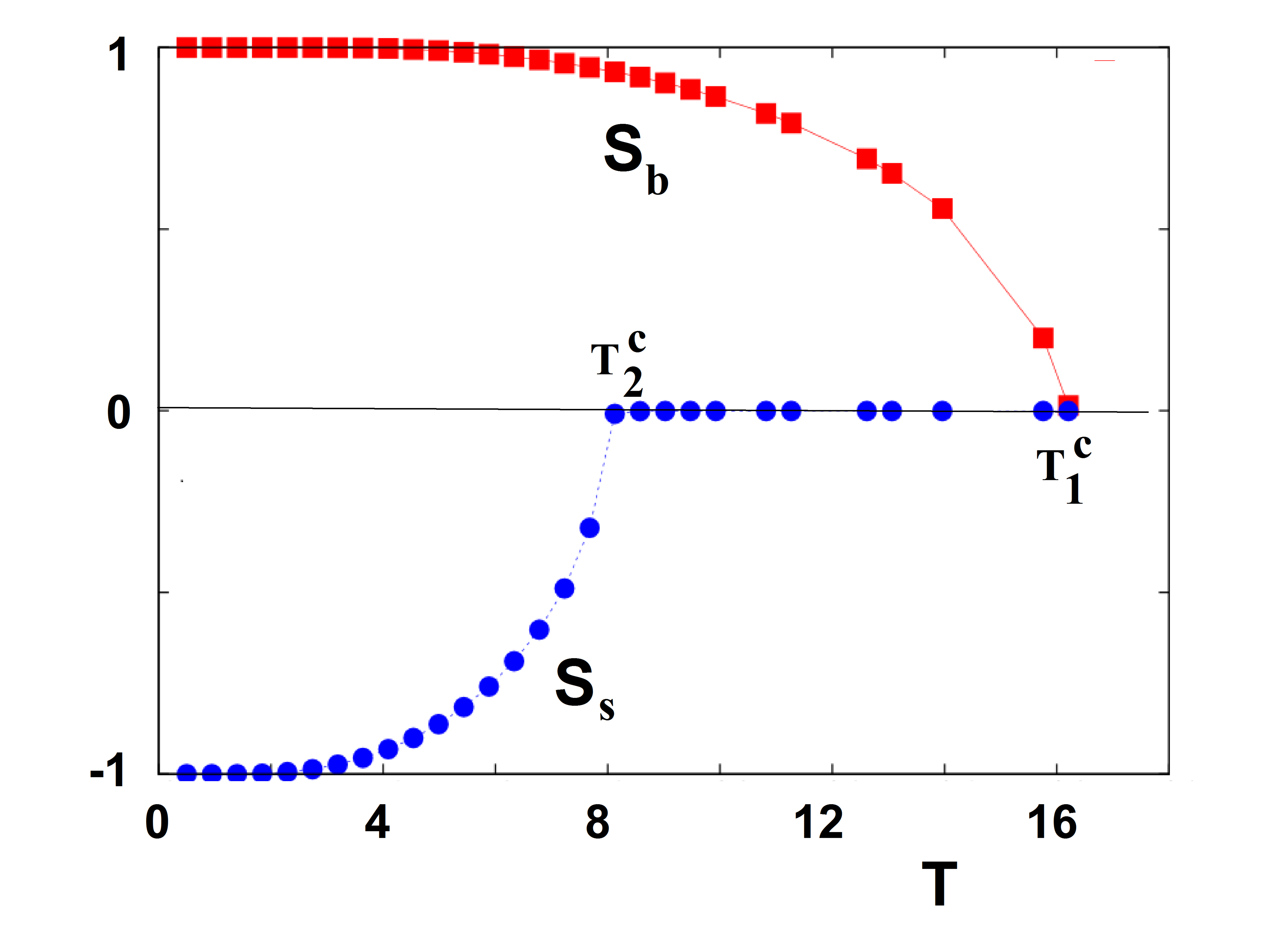}
\vspace{5pt}
\caption{Willing to buy or to sell of two groups, buyers (red squares) $S_b$, and sellers (blue circles) $S_s$, versus market temperature, when there is no inter-group interaction.  Above $T_1^c\simeq 16.1$, group of buyers) does not want to buy, and above $T_2^c\simeq 8.0$ group 2 (group of sellers) does  not want to sell. See text for comments. }
\label{ffig14}
\end{figure}

\noindent Before the interaction between two communities is turned on, the will strength of the group $i$ depends on the intra-group interaction $J_i$. Choosing ($J_1,J_2$), we plot the will of each groups against the market temperature $T$ in Fig. \ref{ffig14}. The values $\langle  s_i \rangle $ in Fig. \ref{ffig14} represents the will degree of group $i$. The higher $T$ the lower $|\langle  s_i \rangle |$ meaning that less people of the group want to sell or buy. At the market temperature $T_i^c$ the group $i$ becomes disordered, namely no market orientation (buying or selling). Due to the difference of $J_1$ and $J_2$, we see that $T_1^c\neq T_2^c$.\\

\begin{figure}[h!]
\vspace{5pt}
\centering
\includegraphics[scale=0.05,angle=0]{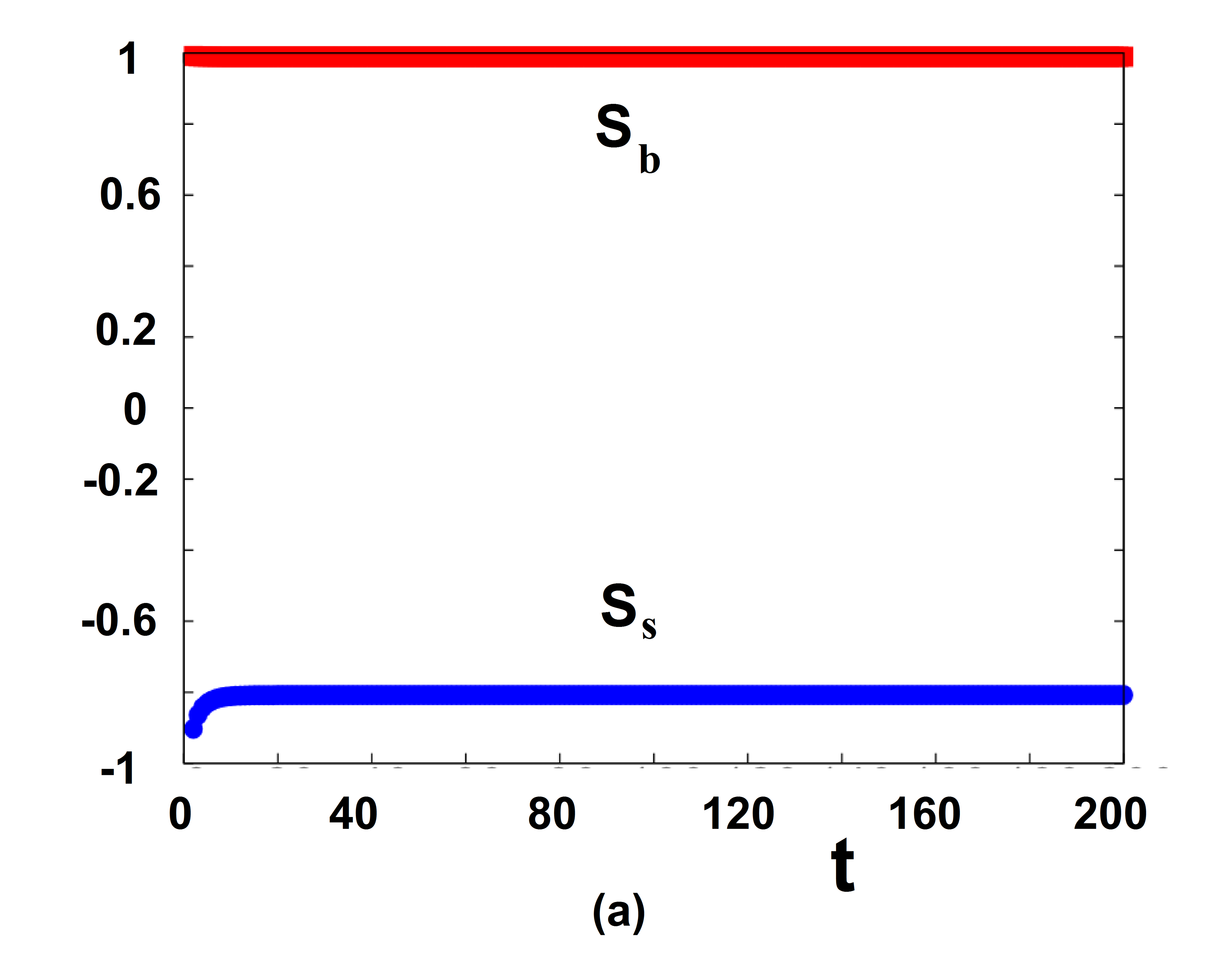}
\includegraphics[scale=0.05,angle=0]{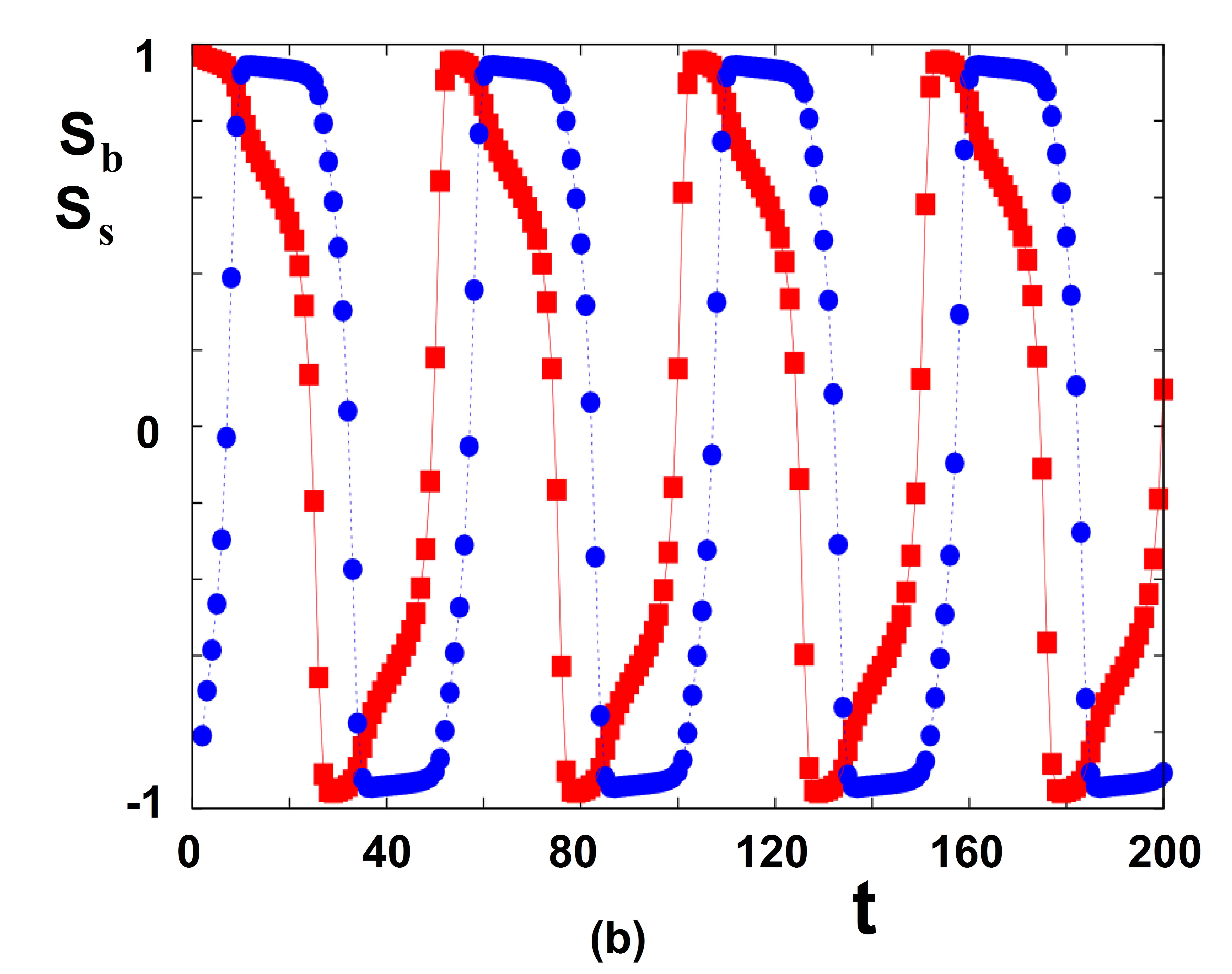}
\includegraphics[scale=0.05,angle=0]{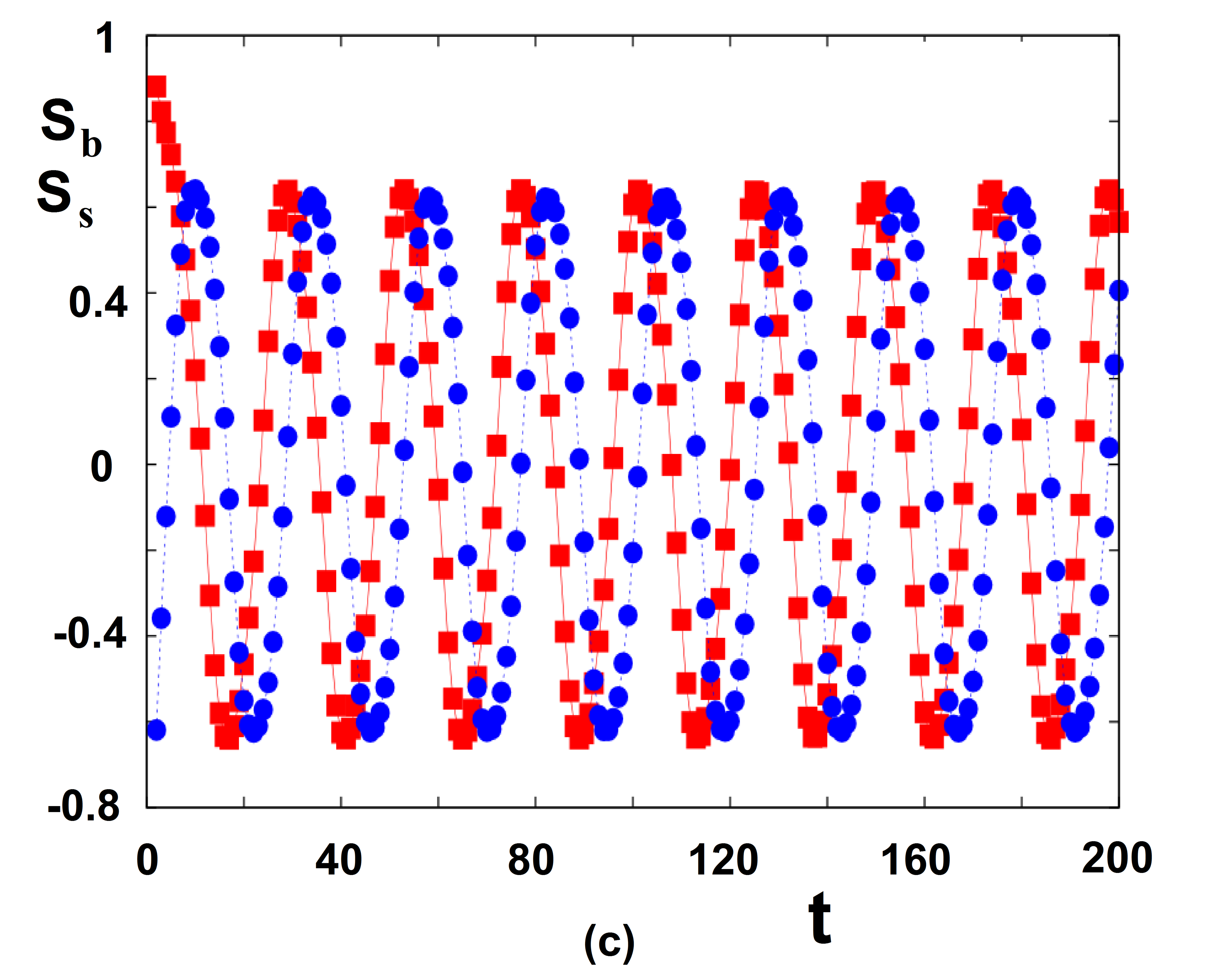}
\includegraphics[scale=0.05,angle=0]{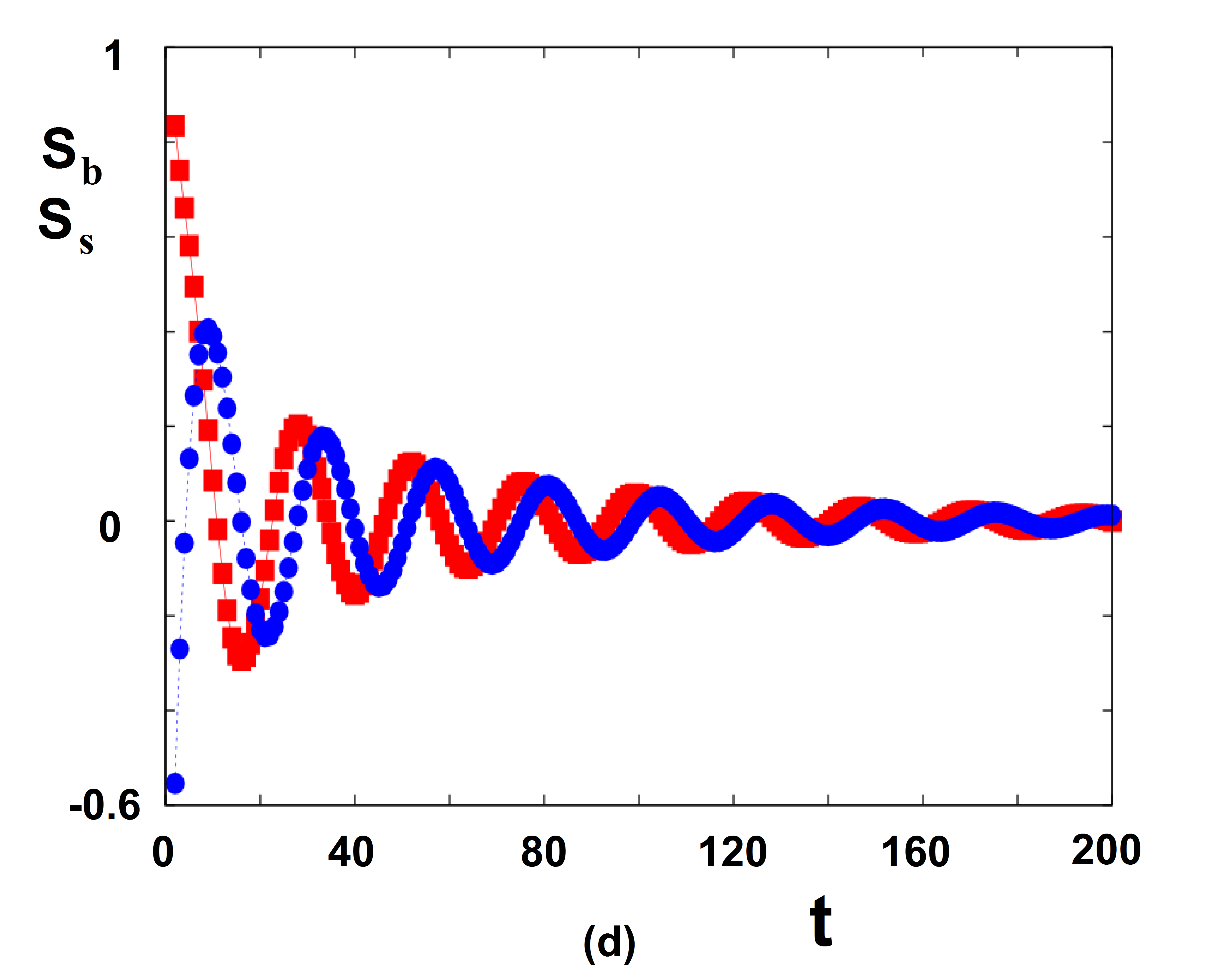}
\vspace{5pt}
\caption{ The percentages of buyers $S_b$ (red) and sellers $S_s$ (blue) at several temperatures for the intra-group and inter-group interactions $J_1=1$, $J_2=0.5$, $K_{12}= 1$, $K_{21}= -0.5$ and $a=5$:  (a) $T=4.98$, (b) $T=6.78$,  (c) $T=10.82$, (d) $T=12.62$.  See text for comments. }
\label{ffig15}
\end{figure}

\noindent We turn on now the interaction between the two groups. For the parameters $J_1=1$, $J_2=0.5$, $K_{12}= 1$, $K_{21}= -0.5$ and $a=5$ used in Fig. \ref{ffig15}, the two groups are both ordered below $T^{c1}\simeq 5.01$ . They both become disordered above $T^{c2}\simeq 12.10$. This is the market clearing point. Between these two temperatures, the systems are dynamically not stable.
We show the time evolutions of $\langle  S_b \rangle $ and $\langle  S_s \rangle $ in Fig. \ref{ffig15} at several market temperatures.  Depending on $T$, the dynamic behaviors of $\langle  S_b \rangle $ and $\langle  S_s \rangle $ are different:
\begin{itemize}
\item At low $T$ namely $T <   T^{c1}$, each group stays in their initial willingness (selling or buying) as shown in Fig. \ref{ffig15}a
\item At a temperature between $T^{c1}$ and $T^{c2}$, the position of each group oscillates as shown in Figs. \ref{ffig15}b and \ref{ffig15}c.  The lower $T$, the larger the period of oscillation (see Ref. \onlinecite{Diep2017})
\item At $T$ near $T^{c2}$, the two groups oscillate at the market opening but tend to the market clearing as seen in Fig. \ref{ffig15}d.
\end{itemize}

\begin{figure}[h!]
\vspace{5pt}
\centering
\includegraphics[scale=0.05,angle=0]{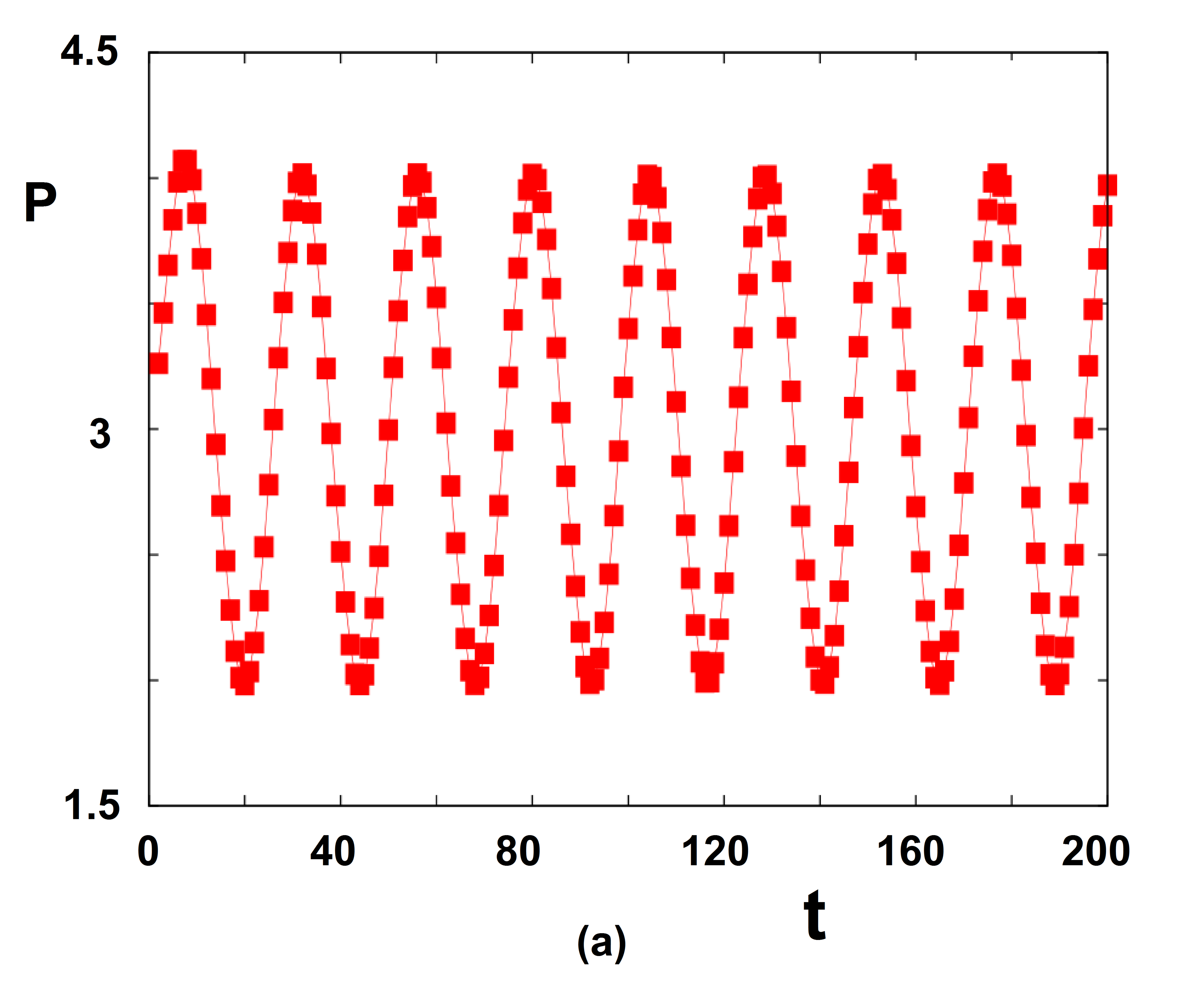}
\includegraphics[scale=0.048,angle=0]{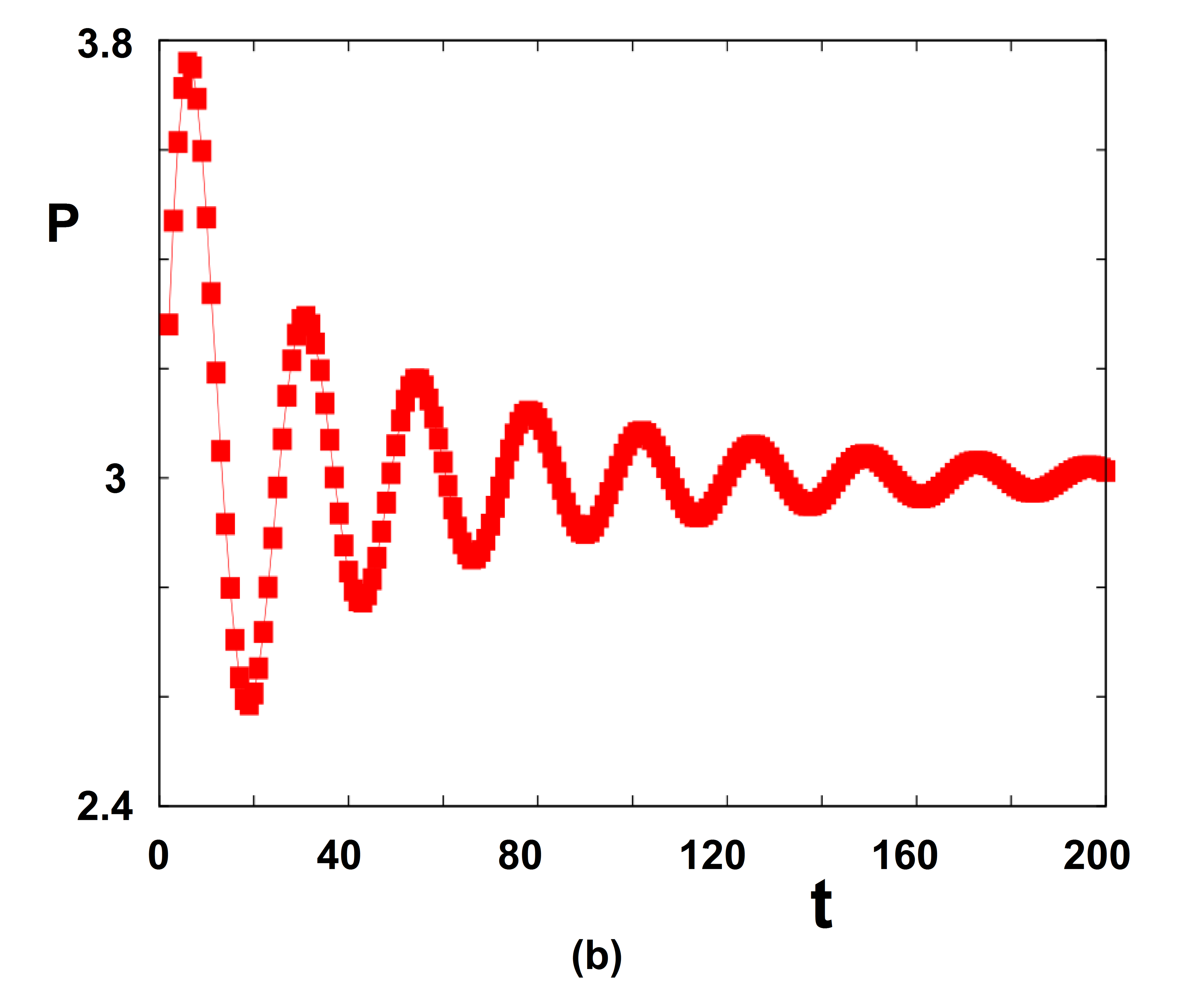}
\vspace{5pt}
\caption{ Price variation versus time for (a) $T=10.82$, (b) $T=12.62$. See text for comments. }
\label{ffig16}
\end{figure}

\noindent We show in Fig. \ref{ffig16} the variation of the price corresponding to the temperatures of Figs. \ref{ffig15}c and \ref{ffig15}d.  The price at the market clearing has been arbitrarily fixed to 3.
The price oscillates for the market temperature between $T^{c1}$ and $T^{c2}$.
For $T  > T^{c2} \simeq 12.10$, the price decays to the market clearing price.\\

\noindent Before closing this section we discuss on the effect of $a$. As seen in Eqs. (\ref{MF1})-(\ref{MF2}), $a$ is in competition with the $j$ and $k$ terms. For a given set of ($j_1,j_2,k_{12},k_{21}$), the value of $a$ determines the qualitative dynamics of the system. We have seen an example for $a=5$ shown above.  We have calculated for other values of $a$. The results show that larger $a$ will enlarge the temperature region of oscillation behavior, and below a value depending on ($j_1,j_2,k_{12},k_{21}$), the oscillation disappears.

\subsection{Summary and discussion}\label{MFdiscussion}

\noindent Since the mean-field theory is based on the principle to replace each individual by a common mean value, fluctuations are neglected in the calculation. This theory cannot explain the nature of the phase transition and overestimates the ordering at low dimensions in statistical physics \cite{DiepSP,DiepTM}. However, the mean-field theory can account for the essential qualitative features which are sufficient for economic issues.  We used this theory for that purpose.\\

\noindent We have seen in the previous section that the Monte Carlo method takes into account these fluctuations which play an important role near the critical economic temperature $T_c$. In mean-field theory, there is no random process as in Monte Carlo simulations: the price at time $t$ depends on the price at $(t-1)$ because the solution is obtained by iteration. So, if we consider only one community composed of sellers and buyers as what we did in Monte Carlo simulations, we will not have random fluctuations of the price with time. We have therefore introduced two interacting communities with no need to have inter-group symmetrical interaction.  As a consequence, we have seen that the price can oscillate below $T_c$ with large amplitudes, namely when the economic agitation is high. Price oscillations are decayed to the market clearing for $T > T_c$.\\

\noindent To conclude this section on the mean-field theory, we emphasize that the model used here is a simplest model aiming at showing the non trivial price oscillation in a region of market temperature. Such a regular oscillation stems certainly from the simplification of the model. We believe however that such a price oscillation bears an important feature of the market reality.

\section{Conclusion}\label{concl}

\noindent In this work, we have studied the variation of the price of a good. This price is determined by the behavior of sellers and buyers. An agent decides to buy or to sell according to a probability which takes into account the influence of his neighbors, the price level, the economic temperature and a specific measure in favor of buying or selling applied during a lapse of time.\\

\noindent We assimilated each agent to a spin possessing several internal states representing the different degrees from the selling desire to the buying desire. The influence of neighbors is expressed by the majority imitation. The price is defined as proportional to the difference between seller and buyer numbers which is a function of time.  The economic environment is expressed by a parameter $T$ which plays the role of the temperature in physics. The higher $T$ the more agitation. The specific measure  $H$ to boost or to lower the price is applied during a lapse of time by the government or an economic organization.\\

\noindent Monte Carlo simulations have been performed on a population of agents with the above-mentioned parameters. The results are interesting, showing in particular strong fluctuations of the price in the critical region of $T$.
There are two stylized facts that the simulation reproduces: random variation of the price as observed in finance, and volatility clustering due to the autocorrelation related to the initial condition at the opening of the market or due to the long autocorrelation in the critical region (the so-called critical slowing-down).\\

\noindent Another striking finding observed in the simulation is the persistent effect of a temporary shock $H$ such as a boosting measure applied during a short lapse of time.  This phenomenon is observed only in the region of economic turbulence (near $T_c$) with $H$ strong enough, above a critical value $H_c$. The boosting effect lasts long time after the removal of the measure. These results suggest that the government and/or an economic organization has to choose the right moment to intervene in order to modify the market tendency for a long time.\\

\noindent It is interesting in a future work to calculate the Hurst exponent \cite{Hurst1,Hurst2} used as a measure of long-term memory of time series \cite{Torsten,Quian}. This exponent relates to the autocorrelations of the time series, and the rate at which these autocorrelations decrease as the lag between pairs of values increases.\\

\noindent The second part of the paper deals with the price variation using the mean-field theory. Unlike the Monte Carlo model, we suppose here that the sellers and buyers belong to two distinct communities. Interesting results on the price oscillation are found when we break the symmetry of the inter-group interactions, namely $J_1\neq J_2$ and $K_{12}\neq K_{21}$.  We believe that this work paves the way for future more realistic models using statistical physics approach.

\end{document}